\documentclass[graybox]{svmult}

\usepackage{type1cm}        
\usepackage{makeidx}         
\usepackage{graphicx}        
\usepackage{multicol}        
\usepackage[bottom]{footmisc}

\usepackage{newtxtext}       %
\usepackage{newtxmath}       

\usepackage{bm}
\usepackage{booktabs}
\usepackage{cite}

\makeindex             

\begin{document}
\title*{Computational thermal multi-phase flow for metal additive manufacturing}

\author{Jinhui Yan, Qiming Zhu, Ze Zhao}
\institute{Jinhui Yan \at Department of Civil and Environmental Engineering, University of Illinois at Urbana-Champaign, \email{yjh@illinois.edu}}

%
%
\maketitle
\abstract{ Thermal multi-phase flow simulations are indispensable to understanding the multi-scale and multi-physics phenomena in metal additive manufacturing (AM) processes, yet accurate and robust predictions remain challenging. This book chapter summarizes the recent method development at UIUC for simulating thermal multi-phase flows in laser powder bed fusion (LPBF) and directed energy deposition (DED) processes. Two main method developments are discussed. The first is a mixed interface-capturing/interface-tracking computational framework aiming to explicitly treat the gas-metal interface without mesh motion/re-meshing. The second is a physics-based and non-empirical deposit geometry model for DED processes. The proposed framework's accuracy is assessed by thoroughly comparing the simulated results against experimental measurements on various quantities. We also report critical quantities that experiments can not measure to show the predictive capability of the developed methods. }

\section{Introduction} \label{introduction}
Multi-phase flows ubiquitously exist in natural and human-made systems. Their numerical simulations have advanced many scientific and technological areas, such as bubble dynamics, propeller design, oil refining, and chemical reaction optimization, to name a few. In recent years, multi-phase flow simulations are attracting attention from metal additive manufacturing (AM): A type of technology with the potential to reshape various industries because of its superior capability to print complex metallic parts directly from digital models without the constraints of traditional manufacturing technologies~\cite{frazier2014metal,wei2020mechanistic}. In metal AM simulations, thermal multi-phase flow-based models, which solve coupled multi-phase Navier-Stokes and thermodynamics equations with phase transitions, are widely deemed the state-of-the-art predictive tools with the highest fidelity. They complement the expensive experiments, such as in-situ high-speed, high-energy x-ray imaging, to reveal the multi-scale and multi-physical phenomena and derive the process-structure-property-performance relationship in metal AM. Many researchers have proposed various methods in this direction. For example, Lawrence Livermore National Laboratory developed a thermal-fluid solver using the arbitrary-Lagrangian Eulerian (ALE) technique~\cite{ALE3D0,ALE3D1, ALE3D2, ALE3D3}; Yan's group at National University of Singapore developed a set of volume-of-fluid (VoF) based thermal-fluid models to simulate metal AM problems, including directed energy deposition (DED) and multi-layer and multi-track laser powder bed fusion (LPBF) processes~\cite{yan2017multi,yan2018meso, wentao0,wentao1, wentao2}; Panwisawas et al. also used a VoF approach by using OpenFOAM to analyze the inter-layer and inter-track void formation~\cite{panwisawas2017mesoscale}; Lin et al. developed a control-volume finite element approach to simulate DED and LPBF processes~\cite{lin2019numerical,Lin2020}. Li et al. developed a thermal-fluid model by combining the level set method and Lagrangian particle tracking to investigate powder-gas interaction in LPBF processes~\cite{wenda0}. 

Despite the progress that has been made, thermal multi-phase flow simulations for metal AM applications still impose tremendous challenges on numerical methods. The first challenge is how to treat the gas-metal interface, where AM physics, such as phase transitions and laser-material interaction, mainly occur. There are two types of approaches to handle material interface evolution in multi-phase flows. This first option is interface-tracking, including arbitrary Lagrangian-Eulerian (ALE)~\cite{Hughes81a}, front-tracking~\cite{unverdi1992front}, boundary-integral~\cite{best1993formation}, and space-time~\cite{Tezduyar92a}. The material interface evolution in interface-tracking approaches is explicitly represented by a deforming and compatible mesh that moves with the interface. These approaches possess high accuracy per degree of freedom (DoF) and have been applied to many free-surface flow problems~\cite{guler1999parallel}. However, mesh motion and even re-meshing are often required if the interface undergoes large deformations or singular topological changes, which turn out to be very common in metal AM applications even without considering powders. Another option is interface-capturing, including level set~\cite{sussman1994level, osher1988fronts}, front-capturing~\cite{shirani2005interface}, volume-of-fluid (VOF)~\cite{hirt1981volume}, phase field~\cite{jacqmin1999calculation, liu2014thesis}, and diffuse-interface methods~\cite{yue2004diffuse,amaya2010single}. In interface-capturing approaches, an auxiliary field is defined in an Eulerian domain to represent the interface implicitly. The evolution of the interface is governed by an additional scalar partial differential equation (PDE).  Because the interface evolution is embedded in the PDE, these approaches can automatically handle topological interface changes without requiring mesh motion or re-meshing procedures. Interface-capturing approaches have been widely applied to a wide range of interfacial problems, including bubble dynamics~\cite{Jansen2005,tripathi2015dynamics,van2005numerical}, jet atomization~\cite{gimenez2016surface}, and free-surface flows~\cite{calderer2015residual,zhu2019}. However, interface-capturing approaches need higher mesh resolution around the interface to compensate for their lower accuracy. Furthermore, in metal AM applications, an implicit representation of the gas-metal interface imposes technical burdens to handle the laser-material interaction, such as the multiple laser reflections on the melt pool interface.

The second challenge is that metal AM processes, compared with other multi-phase flow problems, involve more physics interplay at a wide range of spatiotemporal scales, including thermodynamics, multi-phase melt pool fluid dynamics, phase transitions (e.g., melting, solidification, evaporation, and condensation), laser-metal interaction, and interface topological changes. Besides, the property ratios are larger than these of two-phase flows in ocean engineering. The resulting linear systems have higher condition numbers due to these aspects, introducing convergence issues for partitioned methods and necessitating more robust coupling solution strategies.

This book chapter review two recent method developments from the authors to address the aforementioned challenges in metal AM processes. The first part of the book chapter presents a mixed interface-capture and interface-tracking formulation~\cite{zhu2021mixed}, inspired by the work in~\cite{Tezduyar01a,Tezduyar01a,cruchaga2007numerical}, to explicitly track gas-metal interface without losing mesh flexibility. The reason we call it a mixed formulation is two-fold. (1) We first utilize a level set method to model gas-metal interface evolution (interface capturing). (2) The gas-metal interface is then explicitly re-constructed by triangulating the intersection points between the zero level set and mesh element edges (interface tracking). Such a combination takes full advantage of the level set method's capability of handling topological interface changes and the convenience of explicit interface representation in treating multiple laser reflections. To ensure the level set field's signed distance property, we abandon the Eikonal equation (a PDE)-based re-initialization used in our previous work and develop a purely computational geometry-based approach in this paper. We show that the geometry-based re-initialization approach attains equivalent and sometimes even better performance than the PDE-based counterpart. The method is very efficient, simple to implement on unstructured tetrahedral meshes, and simultaneously constructs a triangulation with an octree data structure for explicitly representing the interface, facilitating the ray-tracing process for multiple laser reflections. We then describe an effective physics-based and non-empirical deposit geometry model without introducing any additional equation to simulate DED processes~\cite{Zhao21a}. The deposit geometry is based on an energy minimization problem subject to a mass conservation constraint. For both models, variational multiscale formulation (VMS) is utilized as a turbulence model to solve the coupled multi-phase Navier-Stokes and thermodynamics equations augmented with melting, solidification, and evaporation models. Although the characteristic length scales of metal AM processes are small, the melt pool flow speed can reach to order of $m/s$, resulting in non-negligible turbulence. Research in~\cite{hong2003vorticity} showed the effects of with and without turbulence model on the predictions of melt pool dimensions. VMS, because of its variational consistency, flexibility, and previous success in multi-phase flow simulations, is a very natural choice for the problems considered in this paper. We also employ density-scaled continuous surface force (CSF) models to handle surface tension, Marangoni stress, recoil pressure, laser flux, and other boundary conditions on the gas-metal interface. The generalized-$\alpha$ method is utilized to integrate the VMS formulation over time. We employ Newton's method to linearize the nonlinear nodal systems at each time step, which leads to a two-stage predictor/multi-corrector algorithm. The resulting linear system is solved in a fully coupled fashion to enhance robustness by using a generalized minimal residual method (GMRES) with a three-level recursive preconditioning technique.

This book chapter is structured as follows. Section~\ref{sec: computation} presents the core computational methods, including level set method, governing equations, geometry-based re-initialization, ray tracing, variational multiscale formulation, time integration, linear solver, preconditioning, mass fixing, and implementation details, for thermal multi-phase flows in general metal AM processes. Section~\ref{label:benchmark} presents three benchmark examples. The first two are classical multi-phase flow problems that are used to compare the PDE-based and geometry-based re-distancing techniques. The third one is to validate the thermal fluid mechanics model by simulating laser spot weld pool flows using a flat free surface. Section~\ref{LPBF} and Section~\ref{DED} describe the application of the developed computational methods to realistic metal AM processes, including LPBF and DED. One should note that the deposit geometry model based on energy minimization is presented in Section~\ref{deposit}. The simulation results are carefully compared with experimental data from Argonne National Laboratory and other available experimental measurements. The quantities that experiments cannot measure are also reported. We summarize the contributions and specify future directions in Section~\ref{sec:conclusion}.


\section{Computational Methods}\label{sec: computation}
\subsection{Level set method}\label{Level_set_sec}

\begin{figure}[!htbp]
	\centering
	{\includegraphics[width=\linewidth]{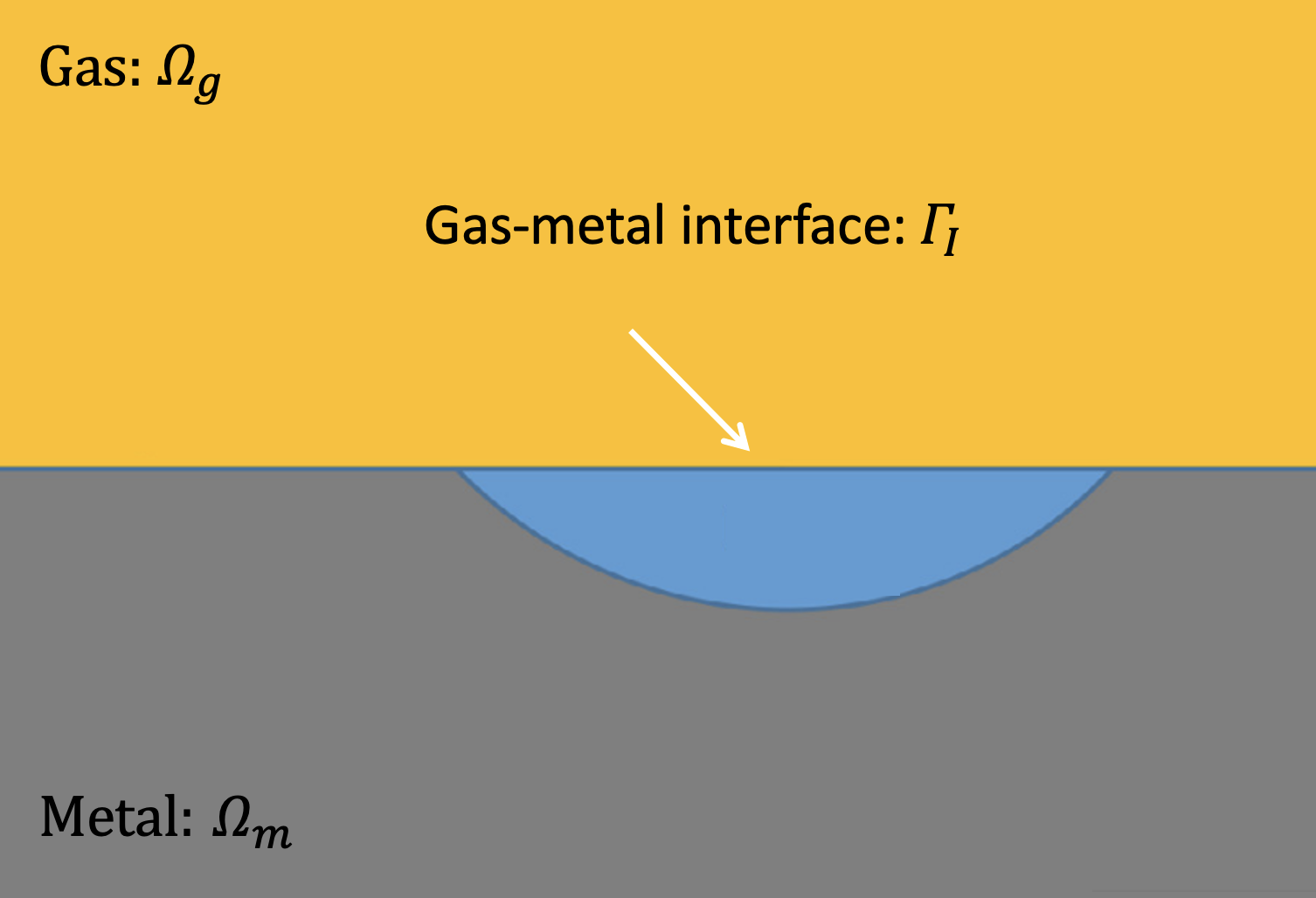}}
	\caption{2-D sketch of the thermofluid system during metal AM processes.}
	\label{fig:metalAM}
\end{figure}

As shown in Fig.~\ref{fig:metalAM}, let $\Omega$ denote the domain of a metal AM problem, consisting of the metal subdomain $\Omega_m$ and gas subdomain $\Omega_g$, and the gas-metal interface $\Gamma_I$ is implicitly represented as
\begin{align}
\label{eq:levelset_phi0}
 \Gamma_I={\bm{x}|\phi(t, \bm{x}) = 0, \forall \bm{x} \in \Omega}
\end{align}
where $\phi (t,\bm{x})$ is a level set field, whose value is the signed distance function from $\bm{x}$ to the gas-metal interface $\Gamma_I$, namely,
\begin{align}
\phi(t,\bm{x}) =
 \begin{cases}
   dist (\bm{x},{\Gamma}_I)  & \text{if $\bm{x} \in {\Omega}_m$ } \\
    -dist (\bm{x},{\Gamma}_I)  & \text{if $\bm{x} \in {\Omega}_g$ }
 \end{cases}
\end{align}
The evolution of $\phi$ is governed by the following convection equation
\begin{align} \label{eq:level_set_govern}
\frac{\partial\phi}{\partial t} + \bm{u}\cdot\nabla\phi = 0 \qquad   
\end{align}
where $\bm{u}$ is the fluid velocity. 
\subsubsection{PDE-based re-initialization}{\label{PDE_redist}}
The signed distance property of level set functions can be polluted by strong convective velocity. One popular re-distancing (or re-initialization) technique is to solve the following pseudo-time dependent Eikonal equation with the constraint on the gas-metal interface. 
\begin{align}
& \frac{\partial \phi_d}{\partial \tilde{t}}+\text{sign}(\phi)(||\nabla \phi_d||-1)=0 & \ \text{in} \ \Omega_m \text{ and } \Omega_g \\
& \phi_d = 0 & \ \text{on} \ {\Gamma}_I\text{ }\\
& \phi_d (\tilde{t} = 0,\bm{x}) = \phi (t, \bm{x}) &  \ \text{in} \ \Omega_m \text{ and } \Omega_g 
\end{align}
where $\tilde{t}$ is the pseudo-time. The pseudo-temporal discretization ($\Delta \tilde{t}$) is scaled by the element length around the interface. The technique was employed in our previous work in conjunction with the variational multi-scale method (VMS) for many multi-phase problems~\cite{yan2018fully, zhu2020immersogeometric,yan2019isogeometric,akkerman2011isogeometric}. The major advantage of this approach is that it only needs to solve a PDE without requiring one particular mesh type. However, an effective pseudo-time integration scheme is necessary, and a linear solver is needed if implemented implicitly. How to choose the pseudo-time step can be tricky for unstructured meshes, and the choice has a significant effect on the accuracy. Besides, lacking an explicit representation of the gas-metal interface still imposes technical burdens on handling the multiple laser reflections in metal AM problems. 

\subsubsection{Geometry-based re-initialization}{\label{GEO_redist}}

We propose a computational geometry-based re-initialization approach specifically designed for metal AM simulations using unstructured meshes. The concept of using geometry for re-initializing level set field can date back to~\cite{ausas2011geometric,strain1999fast} but hasn't been employed in thermal multi-phase flows. As we show below, the approach is simple to implement on unstructured tetrahedral meshes and re-constructs an explicit interface representation from the level set field, which provides significant convenience to handle the multiple laser reflections in metal AM processes.
\begin{figure}[!htbp]
	\centering
	{\includegraphics[width=\linewidth]{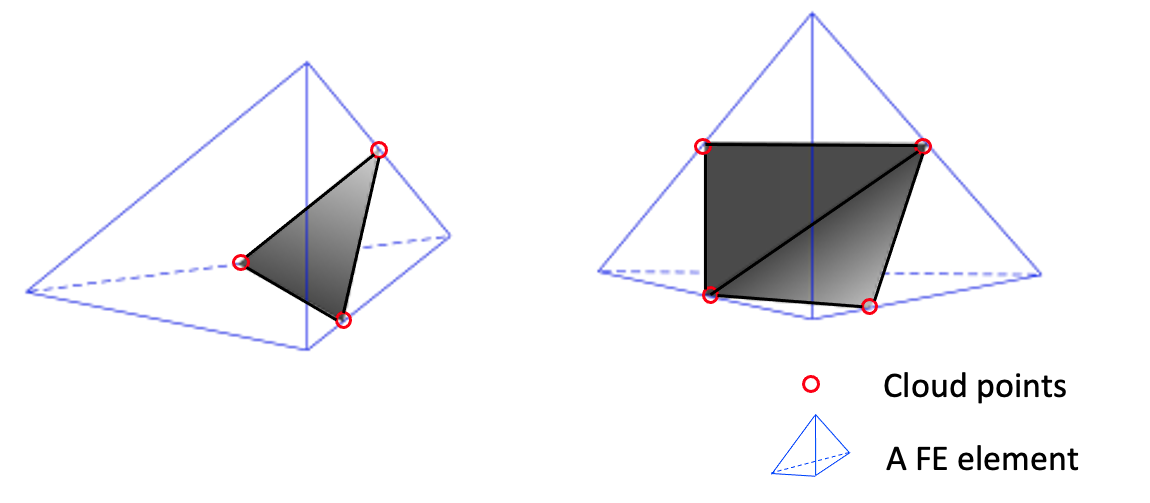}}
	\caption{Element-wise triangulation. First intersection scenario: Three cloud points and one triangle (left). Second intersection scenario: Four cloud points and two triangles (right).}
	\label{fig:tri_phi0_interface}
\end{figure}

\textbf{The first step} of the geometry-based re-initialization is to extract all the intersection points between the gas-metal interface ($\phi = 0$) and every element edge of the mesh. To differentiate from the intersections between laser rays and gas-metal interface, we call these intersections ``cloud points", denoted as a set by $\Xi$. For an element intersected with the gas-metal interface, the cloud points' locations can be obtained by inverting the iso-parametric interpolation. Then, a triangulation of the cloud points is constructed. At first glance, this process seems to require sophisticated algorithms, such as Delaunay triangulation~\cite{liu2013new}. In fact, the triangulation can be performed in an element-wise fashion and suitable for parallel computing. Fig.~\ref{fig:tri_phi0_interface} shows the only possible two intersection scenarios between a tetrahedral element and the gas-metal interface. In the first scenario, the intersection results in three cloud points, and a triangle can be formed by connecting each two of them. In the second scenario, the intersection results in four cloud points, and two triangles can be formed similarly. Then all the triangles are catenated to construct a triangulation that forms an explicit representation of the gas-metal interface.
\begin{figure}[!htbp]
	\centering
	{\includegraphics[width=4.5in]{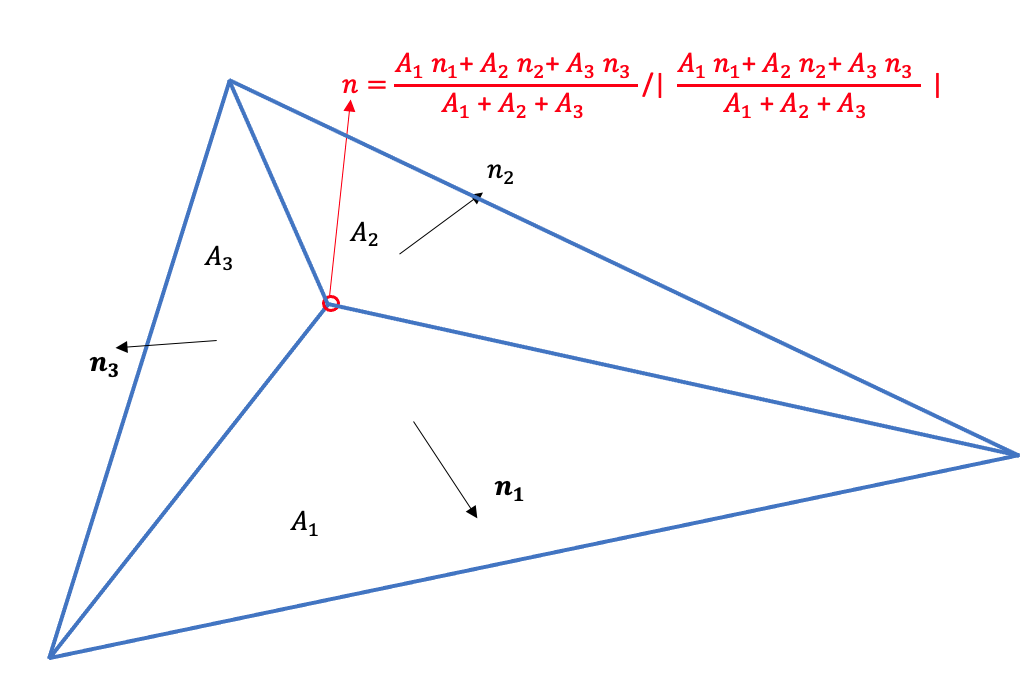}}
	\caption{Unit normal vector definition of a cloud point in the triangulation.}
	\label{fig:normal}
\end{figure}

\textbf{The second step} is to calculate the unit normal vector of the cloud points based on their triangulation. Because the triangulation is only $C_0$ continuous on the cloud points, the normal vector, as shown in Fig.~\ref{fig:normal}, is computed by averaging of the unit normal vector of the triangles (weighted by the areas) associated with this cloud point, namely,  
\begin{align}
    \label{eq:normal_avg}
    \bm{n} = \frac{\sum\limits_i \bm{n}_i A_i}{\sum\limits_i A_i}\LARGE/ \left|\frac{\sum\limits_i \bm{n}_i A_i}{\sum\limits_i A_i}\right|
\end{align}
where $A_i$ and $\bm{n}_i$ are the area and unit normal vector of the triangles. Similar methods can be found in~\cite{jin2005comparison,thurrner1998computing}.

\textbf{The third step} is to restore signed distance for each mesh node by minimizing its normal distance to the gas-metal interface represented by the triangulation of the cloud points. For a mesh node denoted by $\bm{x}_i$, the re-initialized level set $\phi_d (t,\bm{x}_i)$ is defined as 
\begin{align}
    \label{dist_projection}
      \phi_d (t,\bm{x}_i) =    \text{sign}[\phi(\bm{x}_i)] \left | ({\bm{x}_i-\bm{y}_m}) \cdot \bm{n}_m\right |
\end{align}
where $\bm{n}_m$, defined by Eq.~\ref{eq:normal_avg}, is the unit normal vector of the cloud point $\bm{y}_m$, which has the minimal Euclidean distance to $\bm{x}_i$, namely, 
\begin{align}
    \label{k_index_cal}
    m =  \text{arg}\min_{j \in \Xi}\left(|\bm{x}_i-\bm{y}_j|\right)
\end{align}

\begin{figure}[!htbp]
	\centering
	{\includegraphics[width=4.5in]{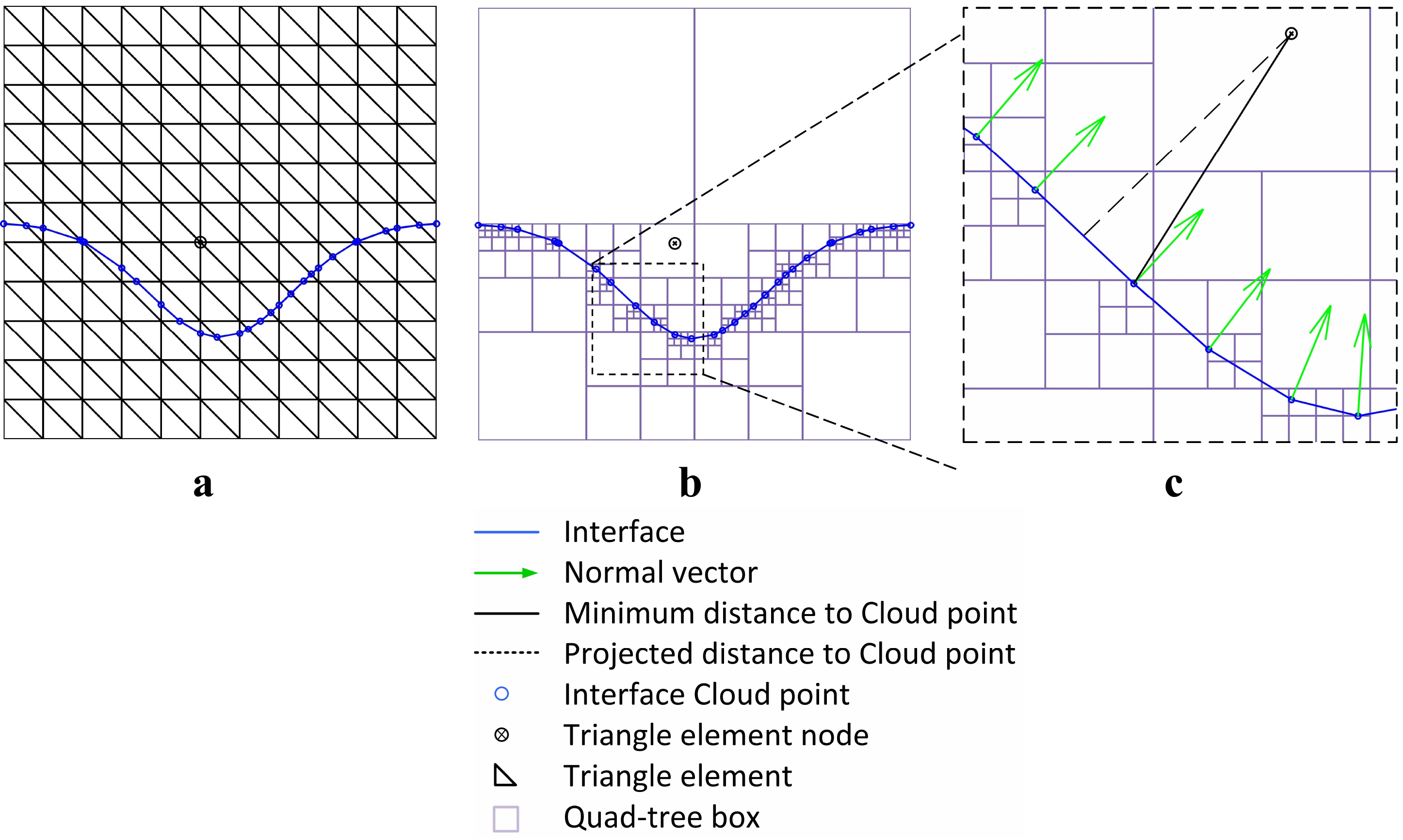}}	
	\caption{A 2D description of the geometry-based re-initialization. (a): Cloud points. (b): Quad-tree structure of the cloud points. (c) Computation of signed distance function.}
	\label{fig:redist_pic}
\end{figure}

One can find the minimal distance and the corresponding cloud point by looping all cloud points. The complexity of this brutal force approach for each mesh node is $O(N)$ if there are $N$ cloud points. The approach can be intractable in metal AM problems, given the large mesh size that also results in a large number of cloud points. To speed up the minimization, we first organize the cloud points $\Xi$ into an octree structure based on the bounding boxes of the domain. Fig.~\ref{fig:redist_pic} (a) and (b) show a 2D presentation of the octree construction process. For a mesh node $\bm{x}_i$, the minimal distance between it and the cloud points are identified by a traversal on the octree with pruning.  The algorithm is described as follows.

\begin{itemize}
\item \textbf{Step 3.1:} Randomly pick a cloud point from the octree, and set the distance between $\bm{x}_i$ and this cloud point to $r_m$ and set the cloud point's index to $m$. Then, start the traversal from the octree root.
\item \textbf{Step 3.2:} For each tree node, calculate the minimum distance between $\bm{x}_i$ and the associated bounding box if $\bm{x}_i$ is outside the current bounding box. If the distance is bigger than $r_m$, skip this path. If not, go to \textbf{step 3.3}.

\item \textbf{Step 3.3:} Repeat the process described in \textbf{step 3.2} for the 8 sub-bounding boxes. If arriving at a leaf of the octree, calculate the minimum distance between $\bm{x}_i$ and the cloud point associated with this leaf. Update the distance and index to the cloud point (in this leaf) which has smaller distance than $r_m$. 

\item \textbf{Step 3.4:} Return $m$ and $r_m$, and calculate re-initialized $\phi_d (t,\bm{x}_i)$ based on Eq.~\ref{dist_projection}.  
\end{itemize}
It is easy to show that the complexity of this approach for each mesh node is only $O(log{N})$. Besides, during the re-initialization, an explicit gas-metal interface, described by the triangulation of cloud points, is constructed, which provides tremendous convenience in the heat laser model, as we show later.

\subsection{Governing equations of thermal multi-phase flows}\label{thermal_fluid_govn}
\subsubsection{Property evaluation}
The thermal multi-phase flows are governed by a unified mathematical model, in which the material properties are phase-dependent. In the model, the level set field is utilized to distinguish the gas phase and metal phase, and the liquid fraction $f_l$ is utilized to distinguish the liquid phase and solid phase in the metal. For a material property $\chi(\phi,f_l)$ (e.g., density, dynamic viscosity, specific heat, and thermal conductivity), it is evaluated by the following linear combination
\begin{align}
\chi(\phi,f_l) = H(\phi)\left[(1-f_l) \chi_s + f_l\chi_l\right]+\left[1-H(\phi)\right]\chi_g
\end{align}
where $\chi_s$, $\chi_l$, and $\chi_g$ are the values of the material property in the solid, liquid, and gas phases, respectively. $H(\phi)$ is a regularized Heaviside function, defined as 
\begin{equation} \label{H_regularized}
 H(\phi)=\left\{
\begin{aligned}
&0 &\phi\  &\leq - &\epsilon \\
&\frac{1}{2} \left(1+\frac{\phi}{\epsilon}+\frac{1}{\pi} sin(\frac{\phi\pi}{\epsilon}) \right) &|\phi| &< &\epsilon \\
&1 &\phi\  &\geq + &\epsilon
\end{aligned}
\right.
\end{equation}
where $\epsilon$ is a numerical gas-metal interface thickness, scaling with the local element size. $f_l$ is a function of temperature $T$, defined as
\begin{align}\label{fl}
 f_l = \begin{cases}
0 & T < T_s \\
\frac{T-T_s}{T_l-T_s} & T_s\leq T \leq T_l \\
1 & T > T_l
\end{cases}
\end{align}
where $T_s$ and $T_l$ are the solidus and liquidus temperatures of the metal material, respectively.
\subsubsection{Navier-Stokes equations of multi-phase flows}

The fluid motion obeys the following Navier-Stokes equations.

\begin{align}\label{continuous-mom}
\rho\left[\frac{\partial\bm{u}}{\partial t}+ \nabla \cdot (\bm{u}\otimes \bm{u})-\bm{g} \right]  -\nabla   \cdot (-p\bm{I} + 2\mu \nabla ^s \bm{u})-\bm{f}_{sf} &=\bm{0} \;\;\;\;\;\;\;\;\;\;\;\;\;\;\;\;\;\;\;\;\;\;\;\;\;\;\;\;\; \text{in} \ \Omega \\
\label{continuous-con}
\nabla  \cdot \bm{u} - m_{e}\bm{n}\cdot \nabla (\frac{1}{\rho}) &= {0} \;\;\;\;\;\;\;\;\;\;\;\;\;\;\;\;\;\;\;\;\;\;\;\;\;\;\;\;\; \text{in} \ \Omega 
\end{align}
where $\bm{u}$ and $p$ are the velocity and pressure fields, $\rho$ is the density, $\mu$ is the dynamic viscosity, $\bm{g}$ is the gravitational acceleration, $\nabla^s$ is the symmetric gradient operator, $\bm{I}$ is a 3 $\times$ 3 identity tensor, and $\bm{f}_{sf}$ represents the interfacial forces that will be defined later. In this model, incompressibility (divergence-free of velocity) still holds in the metal and gas phases individually. However, compressibility is induced at the gas-metal interface due to the local evaporation, which is accounted for by the second term of the continuity equation in Eq.~\ref{continuous-con}, where $\bm{n} = -\frac{\nabla \phi}{|\nabla \phi|}$ is the unit normal vector at the gas-metal interface pointing from metal phase to gas phase. One should note the definition of this normal is different from that of cloud points defined in the previous section. At last, $m_{e}$ is the net evaporation mass flux rate, defined as
\begin{align}
    \label{evap_formula}
    &m_{e} = \zeta P_{sat}(T)\sqrt{\frac{m_{mol}}{2\pi R_{gas}T}}
\end{align}
where the coefficient $\zeta$ accounts for the condensation effect and is set to 0.4 in this paper, $m_{mol}$ is the molar mass of evaporating species, $R_{gas}$ is the gas constant, $P_{sat}$ is the saturation pressure based Clausius-Clapeyron relation, which reads 
\begin{align}
    \label{P_sat_formula}
    &P_{sat}(T) = P_{amb}exp\left[\frac{-L_v m_{mol}}{R_{gas}}(\frac{1}{T}-\frac{1}{T_{evap}})\right]
\end{align}
where $L_v$ is the latent heat of vaporization, $T_{evap}$ is the boiling temperature, $P_{amb}$ is the ambient pressure and set to $101$ KPa in the metal AM simulations here. A complete derivation of Eq.~\ref{continuous-con} can be found in the last author's previous work in~\cite{lin2020conservative}, which adopted a control volume finite element discretization. This evaporation model is also similar to those proposed in~\cite{courtois2013new,courtois2013complete,courtois2014complete,esmaeeli2004computations}. One should note that our approach is based on a continuum model and assumes that the Mach number of the vapor flow is very low, which is valid for the problems considered in the paper. This model cannot handle the extreme situation if the vapor escaping speed is higher than the sound speed because the continuum assumption does not hold in the Knudsen layer and the numerical interface thickness (at the scale of several micrometers) is much larger than the Knudsen layer thickness (at the scale of several mean free path). An alternative laser model that can potentially handle this situation can be found in ~\cite{wang2020evaporation}.

Four types of interfacial forces are modeled through a continuum surface force (CSF) model~\cite{brackbill1992continuum,yokoi2014density} in $\bm{f}_{sf}$, which reads
\begin{align}
\label{f_sum}
\bm{f}_{sf} = \bm{f}_{\sigma}+\bm{f}_{m}+ \bm{f}_{e} + \bm{f}_{r}
\end{align}
where $\bm{f}_{\sigma}$ and $\bm{f}_{m}$ are the surface tension and Marangoni force, defined as
\begin{align}
\label{f_tension}
&\bm{f}_{\sigma} = \delta_{\rho} \sigma \overline{\kappa} \bm{n}\\
\label{f_Marangoni}
&\bm{f}_{m} = \delta_{\rho} \frac{\partial \sigma}{\partial T}\left[\nabla T - \bm{n}(\bm{n}\cdot \nabla T)\right]
\end{align}
where $\sigma =\sigma_0 + \frac{\partial \sigma}{\partial T}(T-T_0)$ is the surface tension coefficient, where $\sigma_0$ is surface tension coefficient at the reference temperature $T_0$, $\frac{\partial \sigma}{\partial T}$ is the Marangoni coefficient, $\delta_{\rho} = (\frac{2\rho}{\rho_m + \rho_g})\frac{\partial H}{\partial \phi}$ is a density-scaled Dirac delta function. $\overline{\kappa}=-\nabla \cdot \bm{n}$ is the mean curvature of the gas-metal interface. $\overline{\kappa}$ calculation needs second-order differential operator, the evaluation of which at quadrature points necessitates a $L_2$ projection if linear tetrahedron elements are employed in spatial discretization. The projection can be avoided if higher-order basis functions, such as isogeometric basis functions~\cite{Cottrell09a,Hughes05a}, are adopted~\cite{yan2019isogeometric}. 

$\bm{f}_{e}$ and $\bm{f}_{r}$ account for the evaporation force and recoil pressure, which reads
\begin{align}
\label{f_evap_formula}
&\bm{f}_{e} = \nabla (\frac{m_{e}^2}{\rho})\\
\label{f_recoil}
&\bm{f}_{r} = -\delta_{\rho} P_{recoil}\bm{n}\\
\end{align}
where $P_{recoil}$ is the recoil pressure, defined as
\begin{align}
    \label{recoil_formula}
    &P_{recoil} = 0.54P_{sat}
\end{align}

\subsubsection{Energy equation}
The temperature field satisfies the following conservation law of enthalpy. 
\begin{align}
\label{eq:temperature}
\rho c_p \frac{\partial T}{\partial t} + \rho c_p \bm{u} \cdot \nabla T + \rho L_m \frac{\partial f_l}{\partial t} + \rho L_m \bm{u} \cdot \nabla f_l = \nabla \cdot (\kappa \nabla T) + Q_{sf}
\end{align}
where $c_p$ is the specific heat, $L_m$ is the latent heat of fusion, $\kappa$ is the thermal conductivity, $Q_{sf}$ is the energy source term handled by a CSF model, which consists of three parts, namely,

\begin{align}
Q_{sf} = Q_{r} + Q_{e} +Q_{l} 
\end{align}
where $Q_{r}$ accounts for the radiative cooling, defined as 
\begin{align}
\label{Q_rad_formula}
&Q_{r} = -\delta_{\rho} \sigma_{sb}\epsilon(T^4 - T_0^4) 
\end{align}
where $\sigma_{sb}$ is the Stefan–Boltzmann constant, $\epsilon$ is the  material emissivity. $Q_{e}$ accounts for the evaporative cooling, defined as
\begin{align}
\label{Q_evap_formula}
&Q_{e} = -\delta_{\rho} L_v m_{e} 
\end{align}
At last, $Q_{l}$ accounts for the heat source, defined as.
\begin{align}
\label{Q_laser_formula}
&Q_{l} = \delta_{\rho} I_{s}
\end{align}
where $I_{s}$ is the equivalent laser ray energy after taking the multiple reflections into consideration. The definition is described in the next section.

\subsection{Ray tracing for multiple laser reflections}\label{laser_ray}
\begin{figure}[!htbp]
	\centering
	{\includegraphics[width=4.5in]{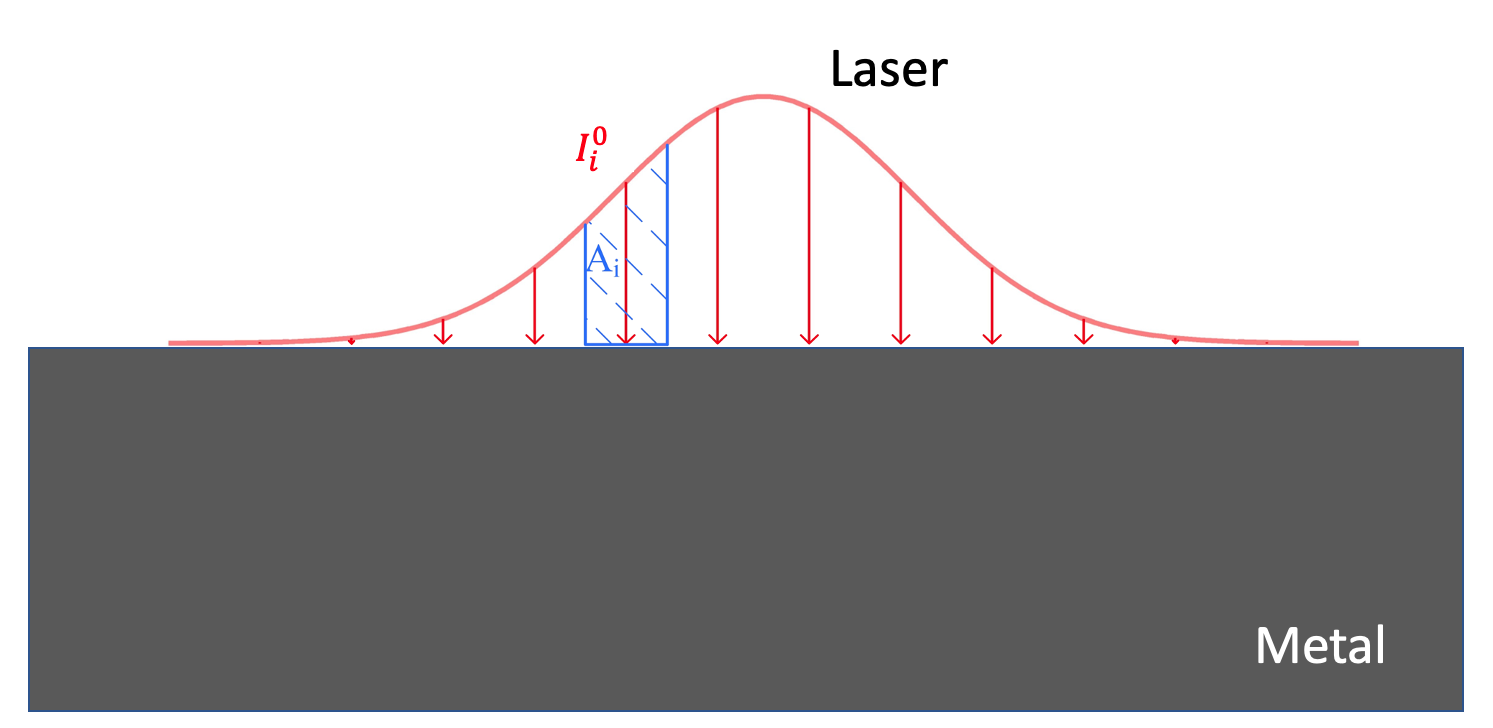}}
	\caption{The laser decomposition into multiple rays.}
	\label{fig:Laser ray}
\end{figure}

Metal AM processes involve violent laser-material interactions. In particular, the multiple laser reflections are vital factors that determine the temperature and melt pool evolution. They must be accounted for carefully to achieve accurate AM process prediction. To this end, a ray-tracing technique is presented~\cite{devesse2015modeling,liu2020new,han2016study,tan2013investigation,yang2018laser}. The laser is uniformly decomposed into $N_{ray}$ rays first. The initial energy of each ray is computed by 
\begin{align}
    \label{laser_power_point_eval}
    &I^0_i = \int_{A_i} I(\bm{x}) d\Gamma
\end{align}
where $i$ in the ray index, ${A_i}$ is the area underneath the ray, and $I(\bm{x})$ is the distribution of original laser, taking a Gaussian profile as
\begin{align}
    \label{laser_gaussian_profile}
    &I(\bm{x})= \frac{2Q\eta}{\pi r_b^2}exp\left[\frac{-2|\bm{x}-\bm{x}_0|^2}{r_b^2}\right]
\end{align}
where $Q$ is the laser power, $r_b$ is the beam radius, $\eta$ is the laser absorption coefficient, $\bm{x}_0$ is the laser center. Fig.~\ref{fig:Laser ray} sketches a 2D description of the decomposition.

\begin{figure}[!htbp]
	\centering
	{\includegraphics[width=4.5in]{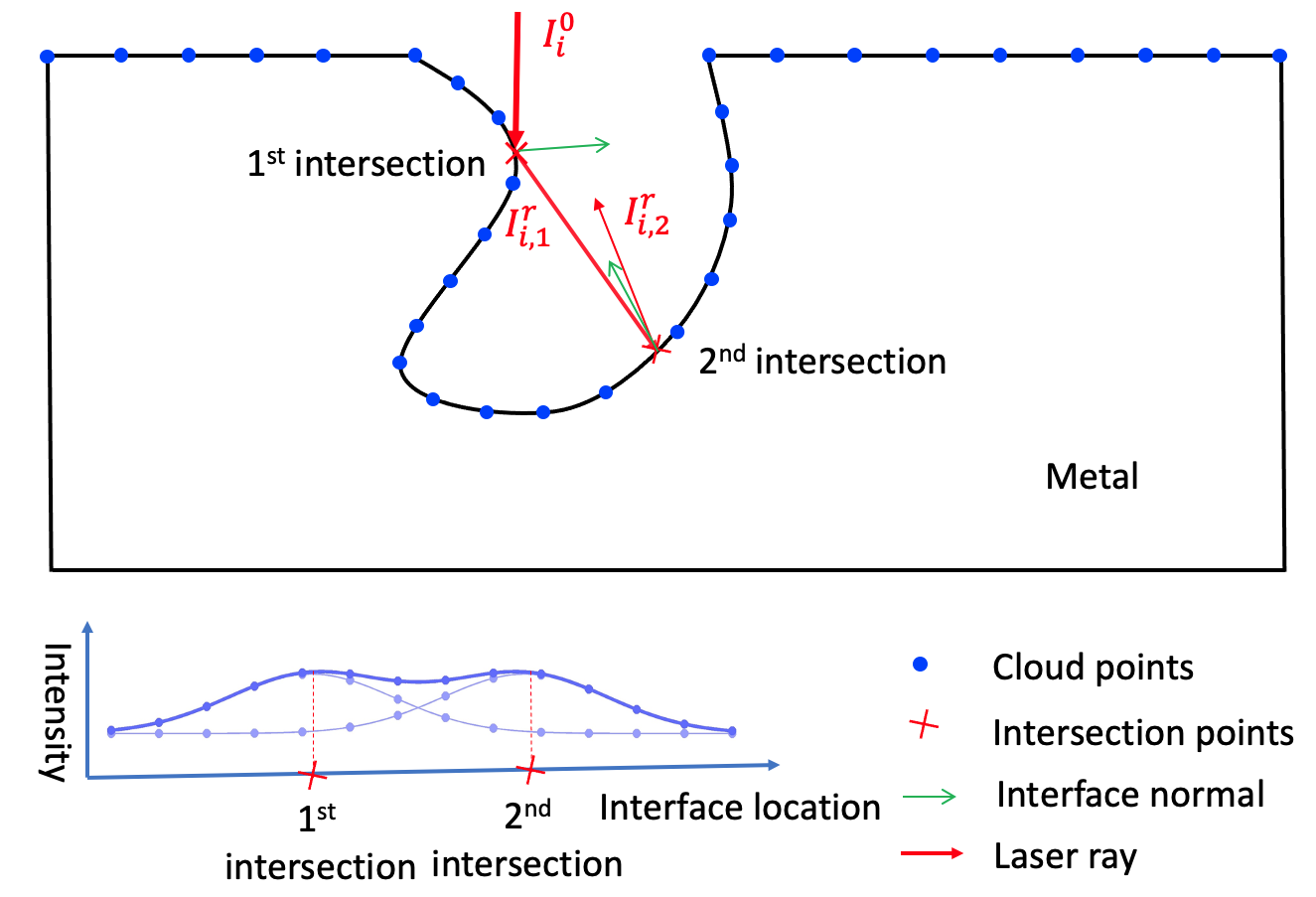}}
	\caption{2D description of the ray tracing and laser model.}
	\label{fig:ray_tracing_laser}
\end{figure}

For each ray, we trace all the intersection points between it and the gas-metal interface during the multiple reflections process. Identifying these intersections can also be accelerated by taking advantage of the octree structure of the cloud points, which has been triangulated in the geometry-based re-initialization stage. The procedure is presented as follows, and a 2D description is shown in Fig.~\ref{fig:ray_tracing_laser}, in which a ray has two intersections with the gas-metal interface. Let $N_i$ denote the number of intersections between the $i$th ray and the gas-metal interface. For the $j$th ($j$ = 1, 2, ..., $N_i$) intersection, the absorbed ray energy, denoted by $I_{ij}^{a}$, and the reflected ray energy, denoted by $I_{ij}^{r}$, are computed as 
\begin{align}
    \label{I_absorb_for}
    &I_{i,j}^{a} = \alpha(\theta)cos(\theta) I_{i,j-1}^r\\
    \label{I_reflect_for}
    &I_{i,j}^{r} = \left[1-\alpha(\theta)\right]cos(\theta) I_{i,j-1}^r
\end{align}
These two recursive relationships imply that the current absorbed and reflected energy come from the ray reflected from the previous intersection. Thus, $I_{i,0}^r = I^0_{i}$. The distribution of absorbed and reflected energy depends on the incident angle $\theta$ and a ray absorption coefficient, also a function of $\theta$, defined as

\begin{align}
    \label{absorp_coeff}
    &\alpha(\theta) = 1 - \frac{1}{2}\left[ \frac{1+(1-\epsilon cos \theta)^2}{1+(1+\epsilon cos \theta)^2} + \frac{\epsilon^2 -2\epsilon cos\theta + 2cos\theta^2}{\epsilon^2 + 2\epsilon cos\theta + 2cos\theta^2}\right]
\end{align}
where $\epsilon$ is a material constant associated the material's electrical conductance. 

For each intersection, the absorbed ray energy is distributed on the gas-metal interface by a Gaussian profile, namely,

\begin{align}
    \label{gauss_for}
    &p_{i,j}(\bm{x}) = \frac{1}{2\pi \beta^2}exp^{- \frac{|\bm{x}-\bm{x}_{i,j}|^2}{2\beta^2}}
\end{align}
where $\bm{x}_{i,j}$ is the coordinates of the intersection, $\beta$ is a length scale that is 3 times of the local element length.
With the above definitions, $I_s$ in Eq.~\ref{Q_laser_formula} is computed by summing of the absorbed energy of all the intersections, namely, 
\begin{align}
    \label{laser_power_interpolation}
    &I_s = \sum\limits_{i=1}^{N_{ray}} \sum\limits_{j=1}^{N_i} I_{i,j}^a p_{i,j}
\end{align}

\subsection{Variational multiscale formulation}
Residual-based variational multi-scale (VMS) formulation is utilized to solve the coupled thermal multi-phase flows equations in Eq.~\ref{eq:level_set_govern}, Eq.~\ref{continuous-mom}, Eq.~\ref{continuous-con}, and Eq.~\ref{eq:temperature}. Let $\mathcal{V}_u$, $\mathcal{V}_p$, $\mathcal{V}_T$, and $\mathcal{V}_{\phi}$ denote the trial function spaces for velocity, pressure, temperature, and level set unknowns, respectively, and $\mathcal{W}_u$, $\mathcal{W}_p$, $\mathcal{W}_T$, and $\mathcal{W}_{\phi}$ denote test function spaces for momentum, continuity, temperature, and level set convection equations, respectively. The semi-discrete formulation based on VMS is stated as follows. Find $\bm{u}\in\mathcal{V}_u$, $p\in\mathcal{V}_p$, $T\in\mathcal{V}_T$, and $\phi \in \mathcal{V}_{\phi}$ such that for all $\bm{w}\in\mathcal{W}_u$, $q\in\mathcal{W}_p$, $s\in\mathcal{W}_T$, and $\eta \in \mathcal{W}_{\phi}$,
\begin{align}\label{ALE-VMS-form}
B_{\text{VMS}}\left(\{\bm{w},q,s,\eta\},\{\bm{u},p,T,\phi\}\right)-F_{\text{VMS}}\left(\{\bm{w},q,s,\eta\}\right)=0
\end{align}
where $B_{\text{VMS}}\left(\{\bm{w},q,s,\eta\},\{\bm{u},p,T,\phi\}\right)$ and $F_{\text{VMS}}\left(\{\bm{w},q,s,\eta\}\right)$ are given as
\begin{align}
\label{B_fluid}
\nonumber B_{\text{VMS}}\left(\{\bm{w},q,s,\eta\},\{\bm{u},p,T,\phi\}\right)&=\int_{\Omega}\bm{w}\cdot\rho\left[\frac{\partial\bm{u}}{\partial t}+ \nabla\cdot (\bm{u}\otimes \bm{u})-\bm{g}\right]~d\Omega\\\nonumber
\nonumber&+\int_{\Omega}\nabla^s\bm{w}: \left(-p \bm{I}+2\mu \nabla^s\bm{u}\right)~d\Omega \\\nonumber
& + \int_{\Omega}q\,\left[\nabla  \cdot\bm{u} - m_{e}\bm{n} \cdot \nabla ({\frac{1}{\rho}})\right]~d\Omega\\\nonumber
\nonumber& -\int_{\Omega}\left( \bm{u}\cdot\nabla  \bm{w}+
\frac{\nabla q}{\rho}\right)\cdot\bm{u}'~d\Omega\\\nonumber
\nonumber&-\int_{\Omega} p'\,\nabla  \cdot\bm{w}~d\Omega +\int_{\Omega} \bm{w}\cdot(\bm{u}'\cdot\nabla  \bm{u})~d\Omega \\\nonumber
&-\int_{\Omega} \frac
{{\nabla \bm{w}}}{\rho}:\left(\bm{u}'\otimes\bm{u}'\right)~d\Omega\\\nonumber
& + \int_{\Omega} s \rho\left[ c_p \left( \frac{\partial T}{\partial t}+\bm{u} \cdot \nabla T\right)+ L_m\left( \frac{\partial f_l}{\partial t} +  \bm{u} \cdot \nabla f_l\right)\right]~d\Omega\\\nonumber
& +\int_{\Omega} \nabla s \cdot \kappa \nabla T  ~d\Omega
- \int_{\Omega} \bm{u} \cdot \nabla s T'~d\Omega\\
&  + \int_{\Omega}{\eta}\left(\frac{\partial{\phi}}{\partial t}+\bm{u}\cdot\nabla  {\phi}\right)~d\Omega - \int_{\Omega} \left( \bm{u}\cdot\nabla  {\eta} \right) \phi' ~d\Omega
\end{align}
\begin{align}
\label{F_fluid}
 F_{\text{VMS}}\left(\{\bm{w},q,s,\eta \right\}) &= \int_{\Omega}\bm{w}\cdot\rho\,\bm{f}_{sf}~d\Omega +\int_{\Omega}s Q_{sf}~d\Omega \\\nonumber
&  + \int_{\Gamma^\text{N}} \bm{w}\cdot\bm{h}_f~d\Gamma  + \int_{\Gamma^\text{T}} {s}{h}_T~d\Gamma
\end{align}
where $\bm{h}_f$ and $h_T$ are the applied fluid traction and heat flux. $\bm{u}'$,  $p'$, and $\phi'$ are the fine-scale velocity, pressure and level set, given as
\begin{align}\label{uprime}
\bm{u}' &= -\tau_\text{M}\left\{\rho\left[\frac{\partial\bm{u}}{\partial t}+ \nabla \cdot (\bm{u}\otimes \bm{u})-\bm{g} \right]  -\nabla   \cdot (-p\bm{I} + 2\mu \nabla ^s \bm{u})-\bm{f}_{sf} \right\}\\
\label{pprime}
p' &= -\tau_\text{C}\,\rho\left[\nabla  \cdot\bm{u} - m_{e}\bm{n} \cdot \nabla ({\frac{1}{\rho}})\right] \\
\label{phiprime}
T' &= -\tau_{T}\left[ \rho c_p(\frac{\partial T}{\partial t} + \bm{u}\cdot\nabla T)+\rho L_m(\frac{\partial f_L}{\partial t} + \bm{u}\cdot\nabla f_L)
-\nabla \cdot(\kappa\nabla T)-Q_{sf}\right]\\
\phi' & = -\tau_{\phi}\left( \frac{\partial{\phi}}{\partial t}+\bm{u}\cdot\nabla  {\phi} \right)
\end{align}
where $\tau_\text{M}$, $\tau_\text{C}$, and $\tau_{\phi}$ are the stabilization parameters, defined as
\begin{align}
\label{momentum-stab-h}
\tau_\text{M} & = \left(\frac{4}{\Delta t^2} +
\frac{4||\bm{u}||^2}{h^2} +
\frac{16\mu^2}{\rho^2 h^4}\right)^{-1/2}\\
\label{cont-stab-h}
\tau_\text{C} &= \frac{h^2}{3\tau_\text{M}}\\
\label{tem-stab-h}
\tau_\text{T} &= \left(\frac{4}{\Delta t^2} +
\frac{4||\bm{u}||^2}{h^2} +
\frac{16\kappa^2}{\rho^2 c_p^2 h^4}\right)^{-1/2}\\
\label{phi-stab-h}
\tau_{\phi} &= \left(\frac{4}{\Delta t^2} + \frac{4||\bm{u}||^2}{h^2} \right)^{-1/2}
\end{align}
where $h$ is the minimum edge length of a tetrahedron element. The above formulation features an extension of the residual-based VMS of single-phase turbulent flows, first introduced in~\cite{Bazilevs07b}, to thermal multi-phase flow problems. The terms in Eq.~\ref{B_fluid} without involving fine-scale quantities are the Galerkin formulations of Navier--Stokes, temperature, and level set convection equations, respectively. The rest can be interpreted as a stabilized method for convection-dominated problems or a large eddy simulation (LES) turbulence model. VMS and its variants, such as ALE-VMS~\cite{Bazilevs08a, Takizawa11n, Bazilevs12a, Bazilevs13a, Bazilevs13b, Bazilevs15b, Bazilevs19a, Masud3} and ST-VMS~\cite{Takizawa11m, Takizawa12e,Takizawa13b, Takizawa14g,Kalro00a,Helgedagsrud18d}, have successfully bean employed as LES models in simulating of a wide range of challenging fluid dynamics and fluid-structure interaction problems. These methods show significant advantages when being deployed to flow problems with moving interfaces and boundaries. Several recent validations and applications include environmental flows~\cite{Zhu20b,Ravensbergen20b,Yan17a,cen2022wall,Bazilevs15a}, wind energy~\cite{Bazilevs10a,Takizawa11a,Takizawa11f,Bazilevs13a,Takizawa13a, Takizawa14c, Takizawa14d, Bazilevs14a,Takizawa15b,Otoguro19b,Ravensbergen20a,Korobenko13b, Bazilevs14c, Bazilevs14d, Korobenko18a, Korobenko18b, Bayram20a,Yan16a, Yan20a,kuraishi2021a,kuraishi2021b,Korobenko17a,bazilevs2016fluid}, tidal energy~\cite{Korobenko20c,Korobenko20b, Korobenko20a,Yan16b,Zhu21a,Yan20a,bazilevs2019computer}, cavitation flows~\cite{Bayram20b, Cen21a}, manufacturing processes~\cite{zhu2021,zhao2022full}, aquatic sports~\cite{Yan15a,Augier14a}, supersonic flows~\cite{Codoni21a}, bio-mechanics~\cite{Terahara19a,Hsu14a,Johnson20b,Takizawa19a,kuraishi2022i,terahara2022ii}, gas turbine~\cite{Otoguro19a,Otoguro18a,Xu17a}, and transportation engineering~\cite{Takizawa15a,Kuraishi14a,Takizawa16i,Kuraishi17a,Kuraishi18a,Kuraishi19b}. 

\subsection{Time integration}\label{Time_integration}
Generalized-$\alpha$ method~\cite{chung1993time,jansen2000generalized} is employed to integrate the VMS formulation in Eq.~\ref{ALE-VMS-form} in time. Without losing generality, let $\bm{R}  = \begin{Bmatrix}
\bm{R}_M,\text{ }  
\bm{R}_C,\text{ } 
\bm{R}_T,\text{ } 
\bm{R}_{\phi} 
\end{Bmatrix}^T$ denote the nodal momentum, continuity, temperature, and level set residuals, $\bm{P}$ denote the nodal pressure unknowns, and $\bm{X} = \begin{Bmatrix}
\bm{u},\text{ }
\bm{T},\text{ }
\bm{\phi}
\end{Bmatrix}^T$ and $\dot{\bm{X}}  = \begin{Bmatrix}
\dot{\bm{u}},\text{ }
\dot{\bm{T}},\text{ }
\dot{\bm{\phi}}
\end{Bmatrix}^T$ denote the nodal velocity, temperature, and level set field unknowns, and their time derivatives. When stepping from $t_n$ to $t_{n+1}$, $\bm{X}$ and $\dot{\bm{X}}$ are linked by the following Newmark-$\beta$ scheme \cite{newmark1959method}
\begin{align}
\label{newmark_vel}
\bm{X}_{n+1}&=\bm{X}_{n}+\Delta t \left[(1-\gamma)\dot{\bm{X}}_{n}+\gamma \dot{\bm{X}}_{n+1}\right]
\end{align}
The reason for separating pressure unknowns from others is that the residual $\bm{R}$ is evaluated at $t_{n+1}$ for pressure but intermediate states between $t_n$ and $t_{n+1}$ for velocity, temperature, and level set. These intermediate states, $\bm{X}_{n+\alpha_f}$ and $\dot{\bm{X}}_{n+\alpha_m}$, are computed as
\begin{align} 
	\label{alpha_udot}
    \dot{\bm{X}}_{n+\alpha_m} &= \alpha_m\dot{\bm{X}}_{n+1} + \left(1-\alpha_m\right)\dot{\bm{X}}_n \\
    \label{alpha_u}
    \bm{X}_{n+\alpha_f} &= \alpha_f\bm{X}_{n+1} + (1-\alpha_f)\bm{X}_{n}
\end{align}
In Eqs.~(\ref{newmark_vel}--\ref{alpha_u}), $\gamma$, $\alpha_m$, and $\alpha_f$ are the parameters of Newmark-$\beta$ and generalized-$\alpha$ methods, chosen based on the unconditional stability and second-order accuracy and requirements~\cite{chung1993time}. 
With the above definitions, the time integration leads to the following nonlinear equations
\begin{equation} \label{NONLINEAR}
\bm{R} (\dot{\bm{X}}_{n+\alpha_m}, \bm{P}_{n+1})  = \begin{Bmatrix}
\bm{R}_M  \\
\bm{R}_C \\
\bm{R}_T\\
\bm{R}_{\phi} 
\end{Bmatrix}_{(\dot{\bm{X}}_{n+\alpha_m}, \bm{P}_{n+1})} =\bm{0}
\end{equation}

The above linear systems are solved in a fully coupled fashion by Newton's method, which results in the following two-stage predictor/multicorrector algorithm with a generalized minimal residual solver (GMRES) enhanced with a recursive preconditioning.

\textbf{Predictor stage}:
\begin{align} 
	\label{alpha_predict_udot}
    \dot{\bm{X}}^0_{n+1} &= \frac{\gamma-1}{\gamma}\dot{\bm{X}}_n \\
    \label{alpha_predict_u}
    \bm{X}^0_{n+1} &= \bm{X}_n \\
    \label{alpha_predict_p}
    P^0_{n+1} &= P_n 
\end{align}
where superscript $0$ indicates the quantities are initial guesses.

\textbf{Multicorrector stage}: Repeat the following procedure until the reduction of the norm of $\bm{R}$ satisfies the tolerance.

\textit{\textbf{Step 1}}. Evaluate intermediate states
\begin{align} 
	\label{alpha_inter_udot}
    \dot{\bm{X}}^l_{n+\alpha_m} &= \alpha_m\dot{\bm{X}}^l_{n+1} + \left(1-\alpha_m\right)\dot{\bm{X}}_n \\ 
    \label{alpha_inter_u}
    \bm{X}^l_{n+\alpha_f} &= \alpha_f\bm{X}^l_{n+1} + \left(1-\alpha_f\right)\bm{X}_n 
\end{align}
where $l$ is a Newton-iteration counter.

\textit{\textbf{Step 2}}. Use the intermediate states to evaluate the right hand side residuals and the corresponding Jacobian matrix, which leads to the following linear systems. 

\begin{equation} \label{LINEAR_SYSTEM}
 \begin{Bmatrix}
 \frac{\partial \bm{R}_l}{\partial \dot{\bm{X}}_{n+1}}\\
 \frac{\partial \bm{R}_l}{\partial \bm{P}_{n+1}}
\end{Bmatrix}  \begin{Bmatrix}
 \Delta \dot{\bm{X}}^l_{n+1}\\
 \Delta  {\bm{P}}^l_{n+1}
\end{Bmatrix} = -\bm{R}_l (\dot{\bm{X}}_{n+\alpha_m}^l, \bm{P}_{n+1}^l)
\end{equation}

The above linear equations are solved to get the increments: $\Delta \dot{\bm{X}}^l_{n+1}$ and $\Delta  {\bm{P}}^l_{n+1}$.

\textit{\textbf{Step 3}}. Correct the solutions with $\Delta \dot{\bm{X}}^l_{n+1}$ and $\Delta {\bm{P}}^l_{n+1}$ as follows
\begin{align}
	\label{alpha_correct_udot}
    \dot{\bm{X}}^{l+1}_{n+1} &= \dot{\bm{X}}^l_{n+1} + \Delta\dot{\bm{X}}^l_{n+1}\\
	\label{alpha_correct_u}
    \bm{X}^{l+1}_{n+1} &= \bm{X}^l_{n+1} + \gamma\Delta t\Delta\dot{\bm{X}}^l_{n+1}\\
    \label{alpha_correct_p}
    \bm{P}^{l+1}_{n+1} &= \bm{P}^l_{n+1} + \Delta \bm{P}^l_{n+1} 
\end{align}

\subsection{Fully-coupled linear solver and recursive preconditioning}
The multicorrector stage requires the solution of a large linear system given by Eq.~\ref{LINEAR_SYSTEM}, which couples different components of the VMS formulation. To increase the formulation's robustness, Eq.~\ref{LINEAR_SYSTEM} is solved by a fully coupled approach, in which the Jacobian matrix is constructed with all the terms in the VMS represented. 
For simplicity, we neglect the time step and iteration counts in the notation and write the Jacobian matrix as
\begin{align}
\label{Jacobian_formula}
\bm{J}&=  \begin{Bmatrix}
 \frac{\partial \bm{R}}{\partial \dot{\bm{X}}}\\
 \frac{\partial \bm{R}}{\partial \bm{P}}
\end{Bmatrix}  = {
\left\{ \begin{array}{cccc}
\frac{\partial \bm{R_M}}{\partial \bm{\dot{u}}} & \frac{\partial \bm{R_M}}{\partial \bm{p}} & \frac{\partial \bm{R_M}}{\partial \bm{\dot{T}}} & \frac{\partial \bm{R_M}}{\partial \bm{\dot{\phi}}}\\
\frac{\partial \bm{R_C}}{\partial \bm{\dot{u}}} & \frac{\partial \bm{R_C}}{\partial \bm{p}} & \frac{\partial \bm{R_C}}{\partial \bm{\dot{T}}} & \frac{\partial \bm{R_C}}{\partial \bm{\dot{\phi}}}\\
\frac{\partial \bm{R_T}}{\partial \bm{\dot{u}}} & \frac{\partial \bm{R_T}}{\partial \bm{p}} & \frac{\partial \bm{R_T}}{\partial \bm{\dot{T}}} & \frac{\partial \bm{R_T}}{\partial \bm{\dot{\phi}}}\\
\frac{\partial \bm{R_{\phi}}}{\partial \bm{\dot{u}}} & \frac{\partial \bm{R_{\phi}}}{\partial \bm{p}} & \frac{\partial \bm{R_{\phi}}}{\partial \bm{\dot{T}}} & \frac{\partial \bm{R_{\phi}}}{\partial \bm{\dot{\phi}}}
\end{array}
\right \}} \end{align}

Due to the complexity of thermal multi-phase flow problems and large property ratios, the condition number of the above Jacobian matrix is very large. Thus, effective preconditioning is necessary. In this paper, we develop a preconditioning strategy constructed in a recursive fashion. To facilitate the derivation, let $\bm{J}_3 = 
\left\{ \begin{array}{ccc}
\frac{\partial \bm{R_M}}{\partial \bm{\dot{u}}} & \frac{\partial \bm{R_M}}{\partial \bm{p}} & \frac{\partial \bm{R_M}}{\partial \bm{\dot{T}}} \\
\frac{\partial \bm{R_C}}{\partial \bm{\dot{u}}} & \frac{\partial \bm{R_C}}{\partial \bm{p}} & \frac{\partial \bm{R_C}}{\partial \bm{\dot{T}}} \\
\frac{\partial \bm{R_T}}{\partial \bm{\dot{u}}} & \frac{\partial \bm{R_T}}{\partial \bm{p}} & \frac{\partial \bm{R_T}}{\partial \bm{\dot{T}}} 
\end{array}\right\}$ denote the velocity-pressure-temperature block in $\bm{J}$, $\bm{J}_2 =
\left\{ \begin{array}{ccc}
\frac{\partial \bm{R_M}}{\partial \bm{\dot{u}}} & \frac{\partial \bm{R_M}}{\partial \bm{p}}  \\
\frac{\partial \bm{R_C}}{\partial \bm{\dot{u}}} & \frac{\partial \bm{R_C}}{\partial \bm{p}} 
\end{array}\right\}$ denote the Navier-Stokes block. Let $\bm{M}$, $\bm{M}_3$, and $\bm{M}_2$ denote the preconditioning matrices for $\bm{J}$, $\bm{J}_3$, and $\bm{J}_2$, respectively. Here $\bm{M}$ is the final preconditioning matrix, which is constructed upon the preconditioning matrix for the level set block $\frac{\partial \bm{R_{\phi}}}{\partial \bm{\dot{\phi}}}$ and $\bm{M}_3$ in the following way
 \begin{align}
\label{preconditioner_vpTphi_block_all}
\bm{M} = \bm{M}_3^* + \bm{M}_{\phi}^* - \bm{M}_{\phi}^*\bm{J}\bm{M}_3^*
\end{align}
where $\bm{M}_3^*$ and $\bm{M}_{\phi}^*$ are defined as
 \begin{align}
\label{M_3^*}
\bm{M}_3^* =  \begin{Bmatrix}
 \bm{M_3} & \bm{O} \\
 \bm{O} &  \bm{O}\\
\end{Bmatrix}
\end{align}
and
 \begin{align}
\label{M_phi}
\bm{M}_{\phi}^* =  \begin{Bmatrix}
 \bm{O} & \bm{O} & \bm{O} & \bm{O}\\
 \bm{O} & \bm{O} & \bm{O} & \bm{O}\\
  \bm{O} & \bm{O} & \bm{O}& \bm{O}\\
 \bm{O} & \bm{O} & \bm{O} &\bm{M}_{\phi}
\end{Bmatrix}
\end{align}
where $\bm{M}_{\phi}$ is a multigrid preconditioning matrix for $\frac{\partial \bm{R_{\phi}}}{\partial \bm{\dot{\phi}}}$. It should be noted these zero blocks in $\bm{M}_3^*$ and $\bm{M}_{\phi}^*$ are designed to ensure that the dimensions of the matrix multiplications in Eq.~\ref{preconditioner_vpTphi_block_all} are consistent.

Similarly, $\bm{M}_3$ is constructed upon the preconditioning matrix for the temperature block $\frac{\partial \bm{R_{T}}}{\partial \bm{\dot{T}}}$ and $\bm{M}_2$ as follows.

\begin{align}
\label{preconditioner_vpT_block}
\bm{M}_3 = \bm{M}_2^* + \bm{M}_{T}^* - \bm{M}_{T}^*\bm{J}_3\bm{M}_2^*
\end{align}
where $\bm{M}_2^*$ and $\bm{M}_{T}^*$ are defined as
\begin{align}
\label{M2}
\bm{M}_2^* =  \begin{Bmatrix}
 \bm{M_2} & \bm{O} \\
 \bm{O} &  \bm{O}\\
\end{Bmatrix}
\end{align}
and 
\begin{align}
\label{MT}
\bm{M}_{T}^* =  
\begin{Bmatrix}
 \bm{O} & \bm{O} & \bm{O} \\
 \bm{O} & \bm{O} & \bm{O} \\
 \bm{O} & \bm{O} & \bm{M}_{T}
\end{Bmatrix}
\end{align}
where $\bm{M}_{T}$ is a multigrid preconditioning for $\frac{\partial \bm{R_{T}}}{\partial \bm{\dot{T}}}$.

Finally, the preconditioning matrix $\bm{M}_{2}$ for the Navier-Stokes block $\bm{J}_{2}$ is defined as

\begin{align}
\label{preconditioner_vp_block}
\bm{M}_2&=
\left[ \begin{array}{cc}
\bm{I} & -(\frac{\partial \bm{R_M}}{\partial \bm{\dot{u}}})^{-1}(\frac{\partial \bm{R_M}}{\partial \bm{p}}) \\
\bm{0} & \bm{I} 
\end{array} 
\right ]
\left[ \begin{array}{cc}
(\frac{\partial \bm{R_M}}{\partial \bm{\dot{u}}})^{-1} & \bm{0} \\
\bm{0} & \bm{S}^{-1}
\end{array} 
\right ]
\left[ \begin{array}{cc}
\bm{I} & \bm{0} \\
-(\frac{\partial \bm{R_C}}{\partial \bm{\dot{u}}})(\frac{\partial \bm{R_M}}{\partial \bm{\dot{u}}})^{-1} & \bm{I} 
\end{array} 
\right ]
\end{align}
where $\bm{S}=(\frac{\partial \bm{R_C}}{\partial \bm{p}}) - (\frac{\partial \bm{R_C}}{\partial \bm{\dot{u}}})(\frac{\partial \bm{R_M}}{\partial \bm{\dot{u}}})^{-1}(\frac{\partial \bm{R_M}}{\partial \bm{p}})$ is the Schur complement. Inverse of matrix is obtained by solving the corresponding linear systems with GMRES. One should note that the construction of $\bm{M}_2$ is different from $\bm{M}_3$ and $\bm{M}$ because of the special structure of Navier-Stokes. The method for $\bm{M}_2$ is similar to the nested preconditioning presented in~\cite{liu2020nested}. The choice of $\bm{M}_2$ is motivated by the fact that pressure serves as a Lagrangian multiplier in the system and $\frac{\partial \bm{R_C}}{\partial \bm{p}}$ is close to zero considering the stabilization terms are relatively small. Another preconditioning choice for $\bm{J}$ is using the inverse of each individual decoupled block for velocity-pressure, temperature, and level set blocks, which has been employed in our previous work in~\cite{yan2018fully} for thermal multi-phase flows. 


\subsection{Mass fixing}\label{mass_conservation_fixing}
Mass conservation is important in multi-phase flow problems \cite{olsson2007conservative,laadhari2010improving}. In the current paper, global mass conservation is preserved by a mass fixing method extended from our previous work to account for the evaporated mass in metal AM problems. The residual of the global metal mass conservation equation of metal is defined as
\begin{align}
     \int_0^T \int_{\Gamma_I} m_{e}(\phi) d\Gamma dt + m_t (\phi)  - m_0
\end{align}
where the first integral is the accumulated evaporated metal mass, $m_0$ is the initial metal mass, $m_t$ are the current metal mass in the domain, defined as
\begin{align}
     m_t (\phi) = \int_{\Omega} \rho_m H(\phi) d\Omega 
\end{align}
where $\rho_m$ is the metal density. To ensure the global mass conversation, the level set field $\phi$, after convection and re-initialization, is perturbed by a global constant $\delta \phi$, the value of which is obtained by solving the following scalar equation, which recovers global mass conservation. 

\begin{align}
    \int_0^T \int_{\Gamma_I} m_{e}(\phi + \delta \phi) d\Gamma dt +  m_t(\phi + \delta \phi) - m_0 = 0
\end{align}
One should note that the above equation is a scalar equation that can be solved efficiently. Besides, since the level set field is globally shifted by $\delta \phi$, it doesn't ruin the signed distance property that is recovered in the re-initialization stage. 
\subsection{Implementation details}
The above mathematical formulation is implemented as follows.
 \begin{itemize}
\item \textbf{Initialization}
\item \textbf{Integrate over time}: For each time step, start with intitial guesses and iterate the following steps until convergence.
   \begin{enumerate}
     \item  Use level set and liquid fraction to get material properties.
     \item  Apply ray-tracing to get the equivalent laser energy in the heat source model.
      \item Linearize RBVMS formulation by Newton's method.
      \item Solve the linear system with the proposed linear solver and preconditioning.
       \item Update the solutions for velocity, pressure, temperature, and level set.
       \item Update liquid fraction.
       \item Update level set field with geometry-based re-initialization.
       \item Update level set field with mass-fixing.
   \end{enumerate}
 \end{itemize}
\section{Benchmark examples}\label{label:benchmark}
We first test the proposed approach's accuracy and modeling capabilities on three benchmark problems.  The first two are representative benchmark problems on level set convection and bubble dynamics, aiming to compare the performance of geometry-based re-initialization and the PDE-based counterpart. The third is a spot welding process, aiming to assess the model's accuracy of temperature prediction.
\subsection{PDE-based re-distancing VS geometry-based re-distancing}\label{label: redistancing}

\begin{figure}[!htbp]
	\centering
	{\includegraphics[width=4.5in]{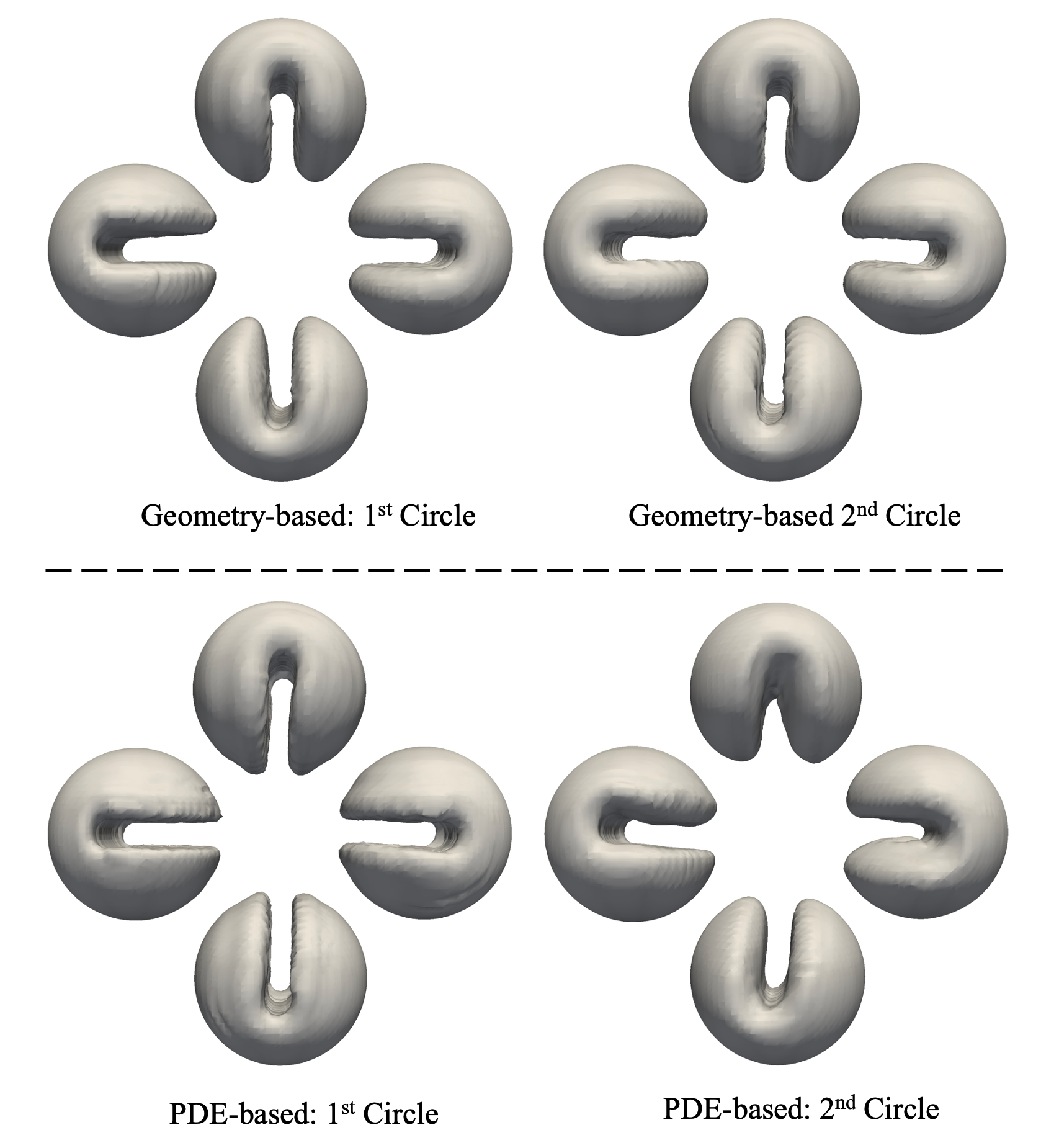}}
	\caption{Sphere shape during two cycles.}
	\label{fig:case1}
\end{figure}

\subsubsection{3D Zalesak problem}\label{case 1}

The 3D Zalesak sphere problem is simulated. This is a pure level set convection problem, in which a slotted sphere with a radius of 0.15 is initialized by level set field at (0.5,0.75,0.5) in a unit cube ($[0,1]\times[0,1]\times[0,1]$). The sphere is convected by the following given velocity

\begin{align}
    \label{case1_vel_for}
    \bm{u} = \frac{\pi}{314}(0.5-x_2,x_1-0.5,0.0)
\end{align}
which rotates the sphere rigidly with angular velocity $\frac{\pi}{314}$ and completes one circle every 628 non-dimensional time units.

 We simulate the problem using both geometry and PDE-based re-initialization approaches, with precisely the same parameters for both. The PDE-based re-initialization is solved with a VMS approach with the generalize-$\alpha$ method for pseudo-time. The details can be found in our previous work in~\cite{yan2016computational,yan2019isogeometric}. An unstructured tetrahedral mesh with uniform element length $h$ =0.01 is used in the simulations. The simulations run for two cycles with $\Delta t$ = 0.5. Fig.~\ref{fig:case1} shows the sphere shape at every quarter during the two cycles. Although both methods preserve the original shape of the slotted sphere well, a noticeable improvement can be observed from the geometry-based case, especially for the second circle. 

\subsubsection{Static bubble}\label{case 2}
 \begin{figure}[!htbp]
	\centering
	{\includegraphics[width=4.5in]{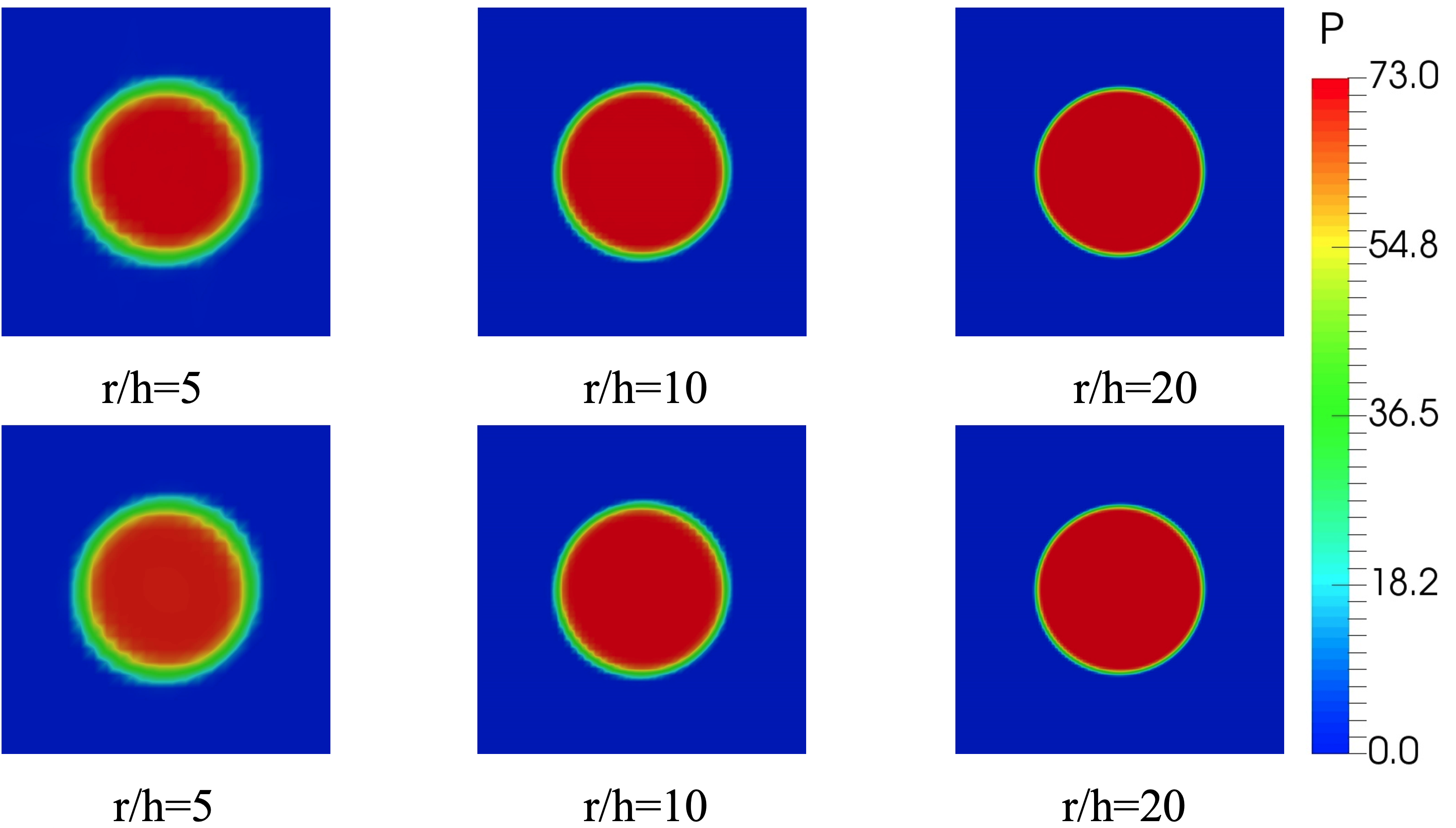}}
	\caption{Pressure contour on three meshes: Geometry-based re-initialization (top). PDE-based re-initialization (bottom).}
	\label{fig:case2_pre}
\end{figure}

\begin{figure}[!htbp]
	\centering
	{\includegraphics[width=4.5in]{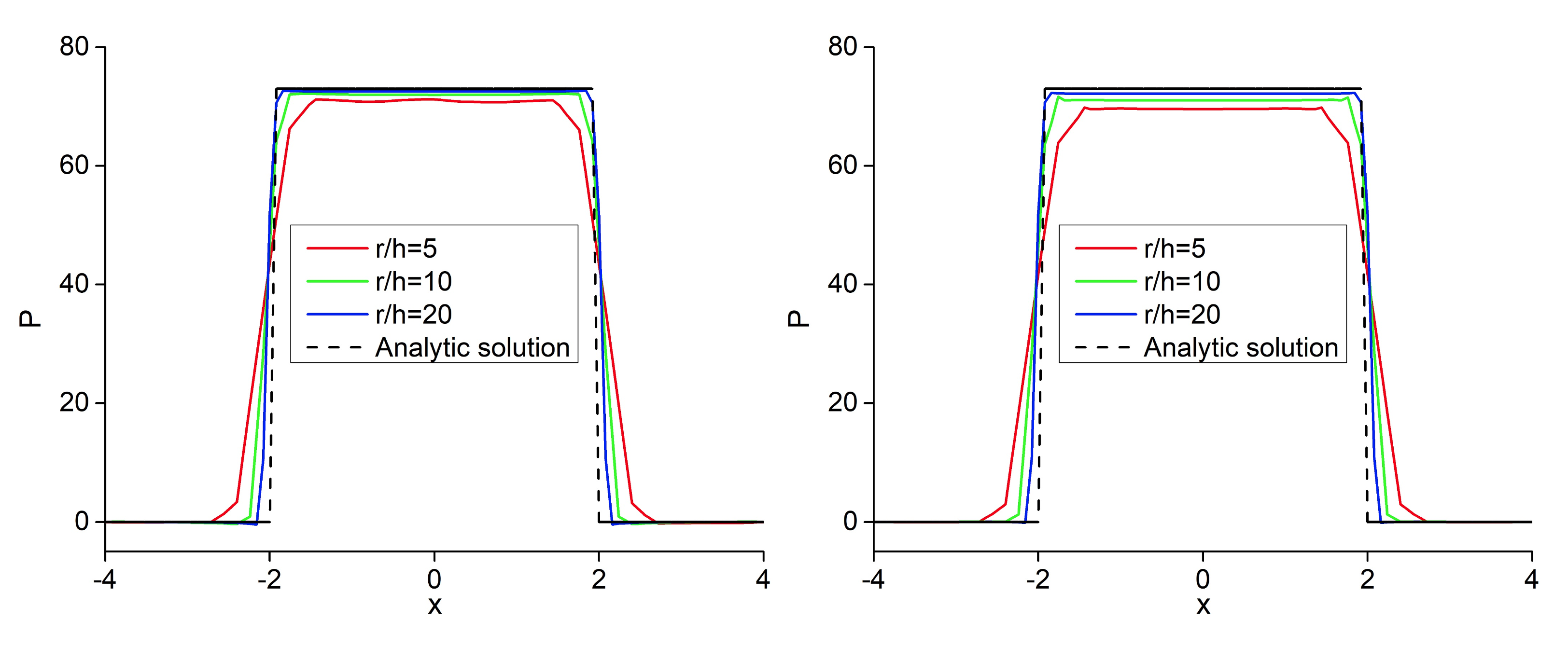}}
	\caption{Pressure along the line from $(-4,0,0)$ to $(4,0,0)$ on three meshes: Geometry-based re-initialization (left). PDE-based re-initialization (right).}
	\label{fig:case2_pre_center_line}
\end{figure}

\begin{figure}[!htbp]
	\centering
	{\includegraphics[width=4.5in]{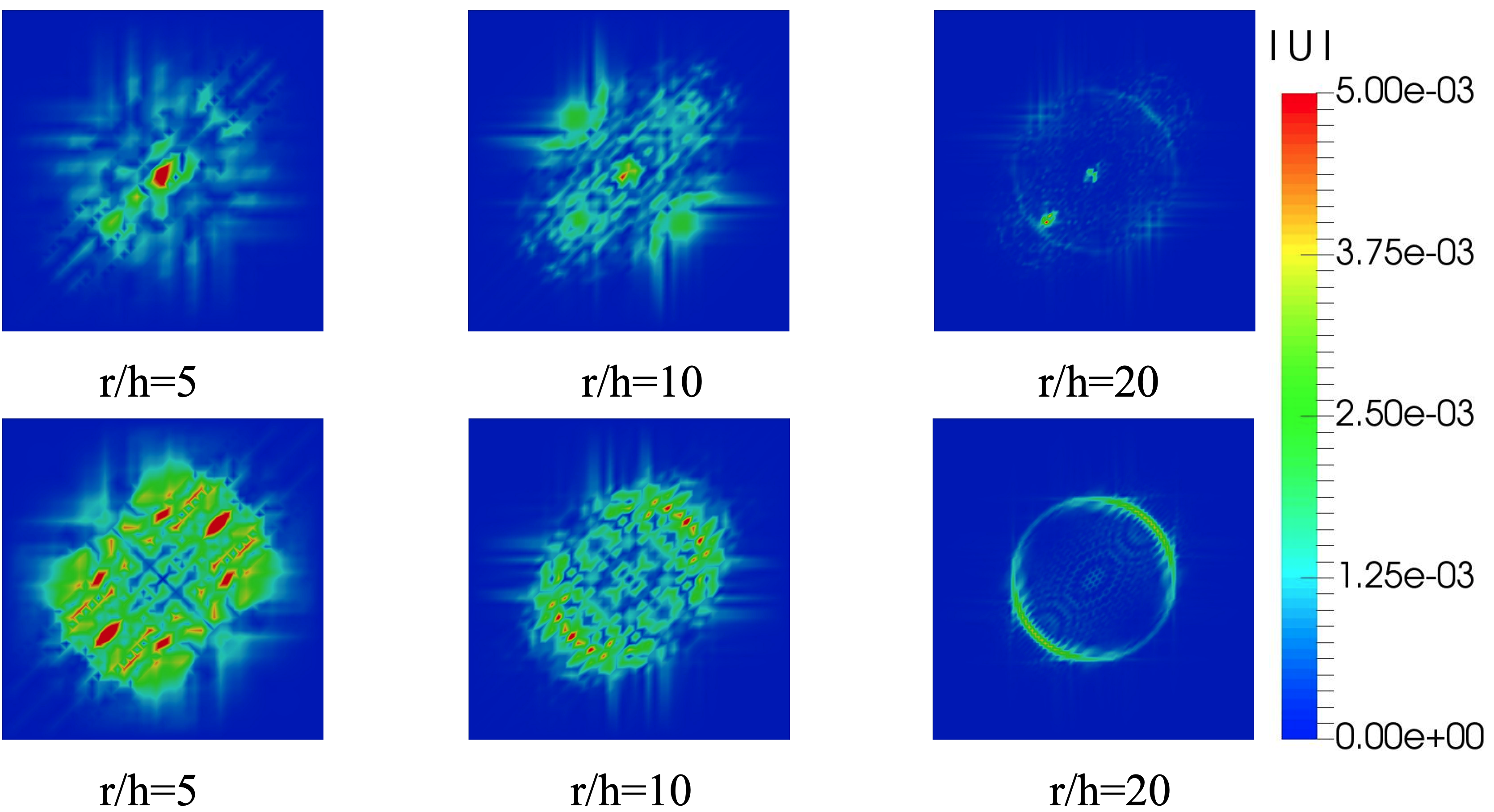}}
	\caption{Velocity magnitude on three meshes: Geometric re-initialization (top). PDE-based re-initialization (bottom).}
	\label{fig:case2_vel}
\end{figure}
We then simulate a static bubble in inviscid liquid without gravity. The exact solutions of this problem are that velocity is zero and pressure difference across the bubble satisfies the following Young-Laplace equation.

\begin{align}
\label{case2_young_laplace_equ}
\triangle p = \sigma \frac{2}{R}
\end{align}
where $R$ is the bubble radius. 

We simulate the problem in a cubic domain ($[-4,4]\times [-4,4] \times [-4,4]$) with unstructured tetrahedral elements. The bubble center is located at (0,0,0). Surface tension coefficient $\sigma$ is set to $73$. The density ratio is $\frac{\rho_l}{\rho_g}=1000$. No-slip boundary condition is employed for all the surfaces. A refinement study with three element lengths, $\frac{R}{h}$=5, 10, and 20, is preformed.

We report the simulated results at the 50th-time step. Fig.~\ref{fig:case2_pre} and Fig.~\ref{fig:case2_pre_center_line} show the pressure contour on a plane cut and along the line from $(-4,0,0)$ to $(4,0,0)$. Both methods produce an accurate prediction on the pressure difference and converge to the exact solution as the mesh is refined. Fig.~\ref{fig:case2_vel} shows the velocity (also called parasite current in the literature) magnitude on the plane of $z=0$. Researchers showed that using the balanced-force CSF model with hard-coded exact mean curvature could reduce the velocity magnitude to zero in machine precision~\cite{francois2006balanced,zhao2020variational,lin2019volume,herrmann2008balanced,montazeri2014balanced}. However, given that it is impossible to obtain analytical mean curvature in real-world problems, we still employ the traditional CSF model and numerically compute the mean curvature by a $L_2$ projection as mentioned above. This CSF model results in non-negligible parasite currents in the domain. Nevertheless, the geometry-based re-initialization produces smaller velocity magnitudes than the PDE-based approach for the three meshes, as seen from Fig.~\ref{fig:case2_vel}.

\subsection{Laser spot weld pool flows}

\begin{table}[!t]
\caption{Material properties.}
 \label{tab:slm_param}
\begin{tabular}{p{4cm}p{4cm}p{2cm}}
\hline\noalign{\smallskip}
Name & Notation (units) & Value  \\
\noalign{\smallskip}\svhline\noalign{\smallskip}
   Gas density & $\rho_a$ $\textcolor{black}{\mathrm{(kg\cdot m^{-3})}}$ & 0.864 \\
   Gas heat capacity & $c_{p,a}$ $\textcolor{black}{\mathrm{(J\cdot kg^{-1}\cdot K^{-1})}}$ & 680 \\
   Gas thermal conductivity & $k_a$ $\textcolor{black}{\mathrm{(W\cdot m^{-1}\cdot K^{-1})}}$ & 0.028 \\
   Metal density & $\rho_l, \rho_s$ $\textcolor{black}{\mathrm{(kg\cdot m^{-3})}}$ & 8100 \\
   Viscosity of liquid metal & $\mu_l$ $\textcolor{black}{\mathrm{(Pa\cdot s)}}$ & 0.006 \\
   Liquid heat capacity & $c_{p,l}$ $\textcolor{black}{\mathrm{(J\cdot kg^{-1}\cdot K^{-1})}}$ & 723.14 \\
   Liquid thermal conductivity & $k_l$ $\textcolor{black}{\mathrm{(W\cdot m^{-1}\cdot K^{-1})}}$ & 22.9 \\
   Solid heat capacity & $c_{p,s}$ $\textcolor{black}{\mathrm{(J\cdot kg^{-1}\cdot K^{-1})}}$ & 627.0 \\
   Solid thermal conductivity & $k_s$ $\textcolor{black}{\mathrm{(W\cdot m^{-1}\cdot K^{-1})}}$ & 22.9 \\
   Liquidus temperature & $T_l$ $\textcolor{black}{\mathrm{(K)}}$ & 1630 \\
   Solidus temperature & $T_s$ $\textcolor{black}{\mathrm{(K)}}$ & 1610 \\
   Latent heat of fusion & $L$ $\textcolor{black}{\mathrm{(J\cdot kg^{-1})}}$ & $2.508\times 10^{5}$ \\ 
\noalign{\smallskip}\hline\noalign{\smallskip}
\end{tabular}
\end{table}

\begin{figure}[!htbp]
 \centering
 \includegraphics[scale=0.27]{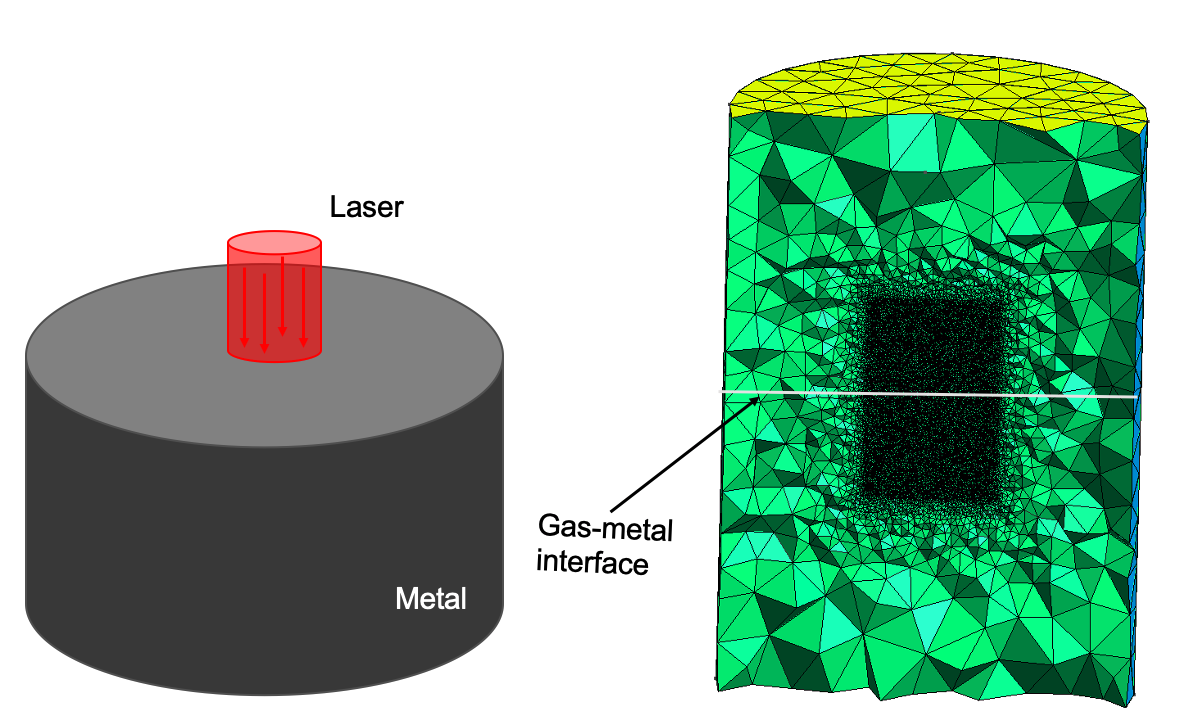}
 \caption{Diagram of laser spot weld pool flow. Left: Problem setup. Right: Mesh.}
 \label{fig:slm_diagram}
\end{figure}
\begin{table}[!t]
 \caption{Parameters of Maragoni coefficient.}
 \label{tab:mgn_exp_weld}
\begin{tabular}{p{4cm}p{4cm}p{2cm}}
\hline\noalign{\smallskip}
Name & Notation (units) & Value  \\
\noalign{\smallskip}\svhline\noalign{\smallskip}
   Pure metal Marangoni coefficient & $\frac{d\sigma}{dT}|_0$ $\textcolor{black}{\mathrm{(N\cdot m^{-1}\cdot K^{-1}) }}$ & $-5 \times 10^{-4} $ \\
   Saturation surface excess & $\tau_s$ $\textcolor{black}{\mathrm{(kmole\cdot m^{-2}) }}$ & $1.3 \times 10^{-8} $\\
   Entropy factor & $k_i\textcolor{black}{(-)}$ & $3.18 \times 10^{-3}$ \\
   Standard heat of absorption & $\Delta H^0$ $\textcolor{black}{\mathrm{(J \cdot kmole^{-1}) }}$ & $-1.66 \times 10^8$ \\
   Partial molar enthalpy of species & $\Delta H^M$ $\textcolor{black}{\mathrm{(J\cdot kmole^{-1}) }}$& $0$ \\
   Sulfur activity & $a_i$ $\textcolor{black}{\mathrm{(\%) }} $ & $0.002, 0.015$\\
\noalign{\smallskip}\hline\noalign{\smallskip}
\end{tabular}
\end{table}

\begin{figure}[!htbp]
 \centering
 \includegraphics[scale=0.8]{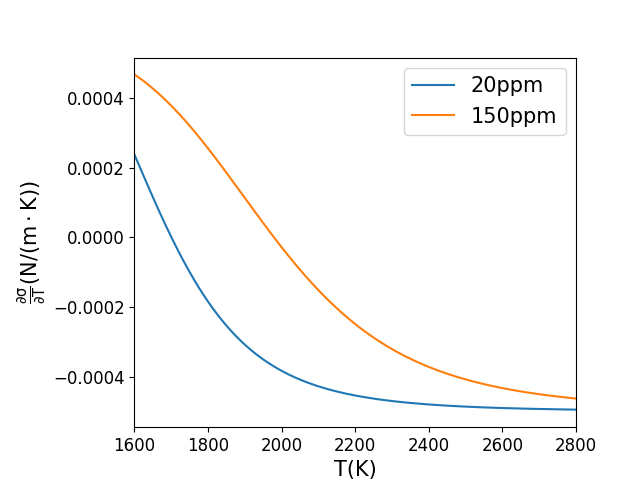}
 \caption{Marangoni coefficient $\frac{\partial \sigma}{\partial T}$ as a function of temperature $T$ for two sulfur activities ($a_i$ = 20 ppm and $a_i$ = 150 ppm).}
 \label{fig:slm_mgn}
\end{figure}

We use the proposed formulation to simulate a laser welding process without material deposition. Fig.~\ref{fig:slm_diagram} (left) shows the problem setup. A bulk of metal based on steel (Fe--S system), with sulfur as the active element, is melted by a stationary heat laser applied on the top surface. The material properties are listed in Table~\ref{tab:slm_param}. The heat laser $q_{in}$ takes the following form.
\begin{align}
q_{in} = 
    \begin{cases}
    \frac{\eta Q}{\pi r_{b}^2} & r \le r_{b} \\
    0 & r > r_{b}
    \end{cases}
\end{align}
where $\eta=0.13$ is the absorptivity, $\textcolor{black}{Q=5200\,\mathrm{W}}$ is the laser power, $\textcolor{black}{r_{b}=1.4\, \mathrm{mm}}$ is the laser radius. The melt pool shape and melt pool fluid dynamics largely depend on the Marangoni coefficient $\frac{\partial \sigma}{\partial T}$. According to the model in~\cite{sahoo1988mgn}, $\frac{\partial \sigma}{\partial T}$ is a function of temperature and sulfur concentration, defined as
\begin{align}\label{Marangoni}
\frac{\partial \sigma}{\partial T}&=\frac{d\sigma}{dT}|_0-R\tau_s ln(1+Ka_i)-\frac{Ka_i}{1+Ka_i}\frac{\tau_s (\Delta H^0 - \Delta H^M)}{T}
\end{align}
where $\frac{d\sigma}{dT}|_0$ is the pure metal Marangoni coefficient, $\tau_s$ is the surface excess at saturation, $R$ is the gas constant. $K = k_i exp(\frac{-\Delta H^0}{RT})$ is the equilibrium constant for segregation, where $k_i$ is the entropy factor, $\Delta H^0$ is the standard heat of absorption, $\Delta H^M$ is the partial molar enthalpy of species mixing in the solution. $a_i$ is the sulfur weight percentage. The values of these parameters are summarized in Table~\ref{fig:slm_diagram}. Two sulfur activities $a_i$ = 0.002\%--wt\,(20 ppm) and $a_i$ = 0.015\%--wt\, (150 ppm) are investigated. The corresponding $\frac{\partial \sigma}{\partial T}$ as a function of temperature is plotted in Fig.~\ref{fig:slm_mgn}.

\begin{figure}[!htbp]
    \centering
    \includegraphics[width=\linewidth]{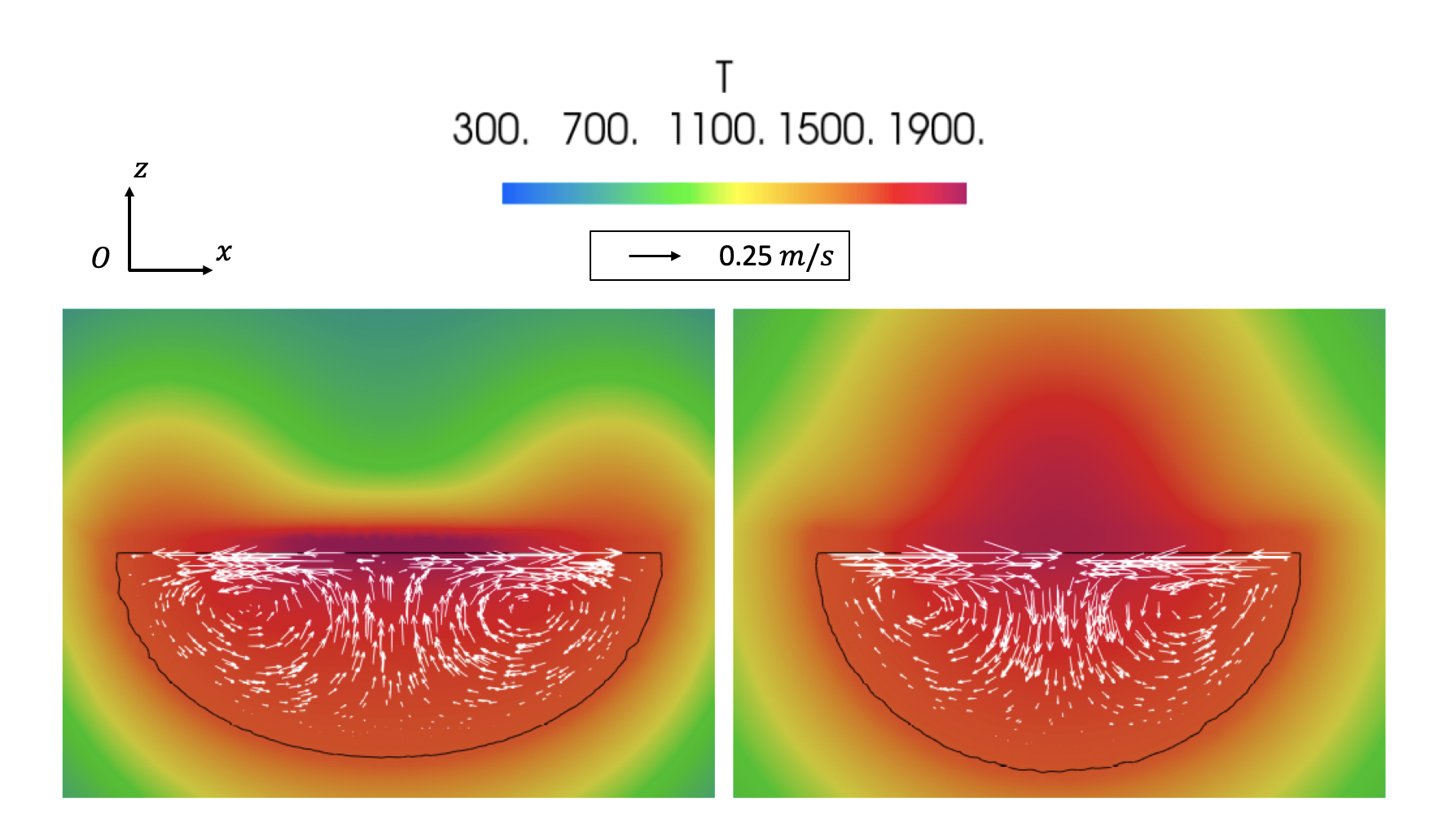}
    \caption{Temperature \textcolor{black}{(unit: K)} and velocity \textcolor{black}{(unit: m/s)} vectors in melt pool (slice view). The black solid line indicates the melt pool boundary. Left: 20 ppm. Right: 150 ppm.}
    \label{fig:slm_field1}
\end{figure}

\begin{figure}
    \centering
    \includegraphics[width=\linewidth]{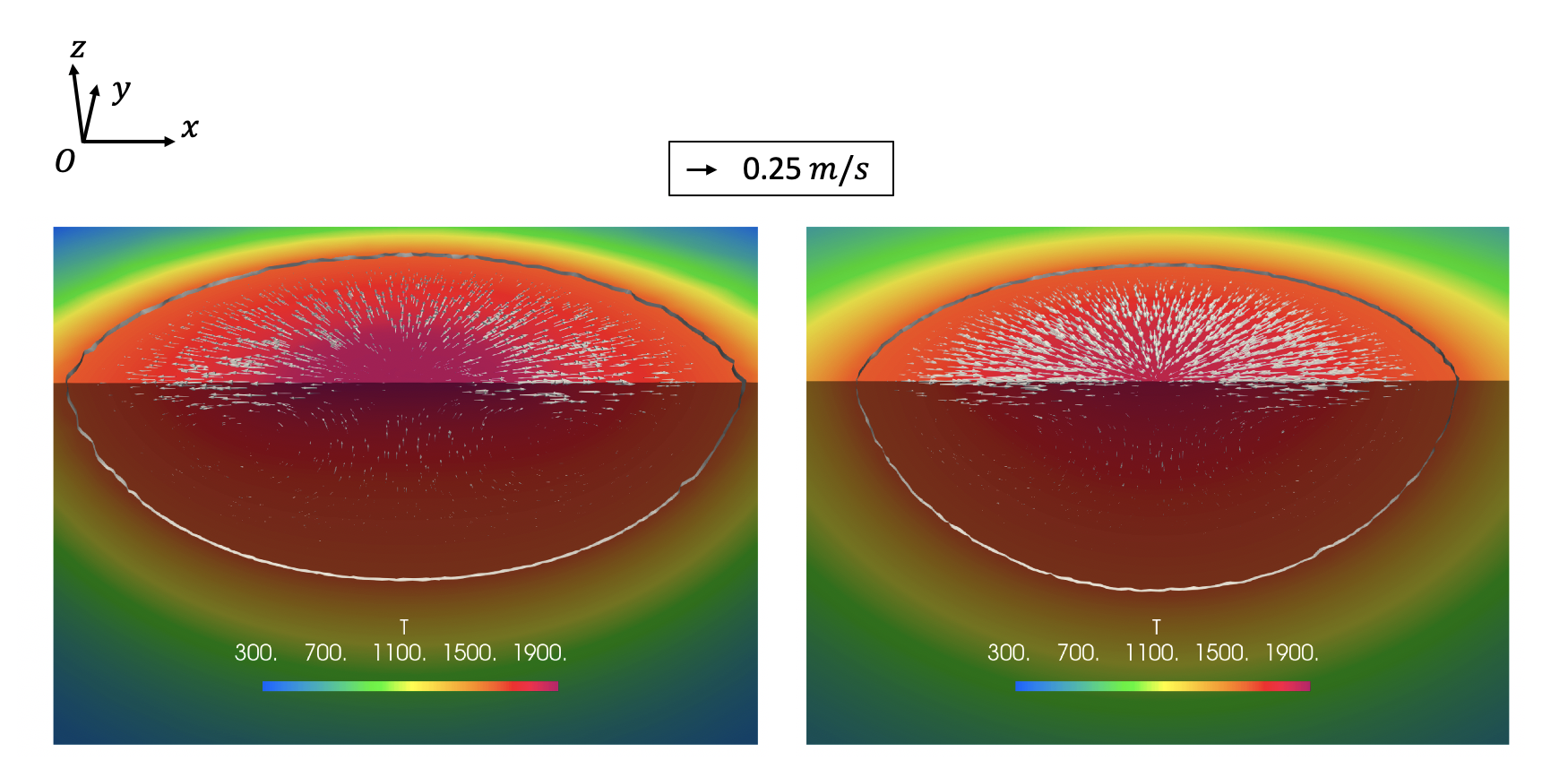}
    \caption{Temperature \textcolor{black}{(unit: K)} and velocity \textcolor{black}{(unit: m/s)} vectors in melt pool (3D view). The black solid line indicates the melt pool boundary. Left: 20 ppm. Right: 150 ppm.}
    \label{fig:slm_field2}
\end{figure}

\begin{figure}[!htbp]
 \centering
 \includegraphics[width=\linewidth]{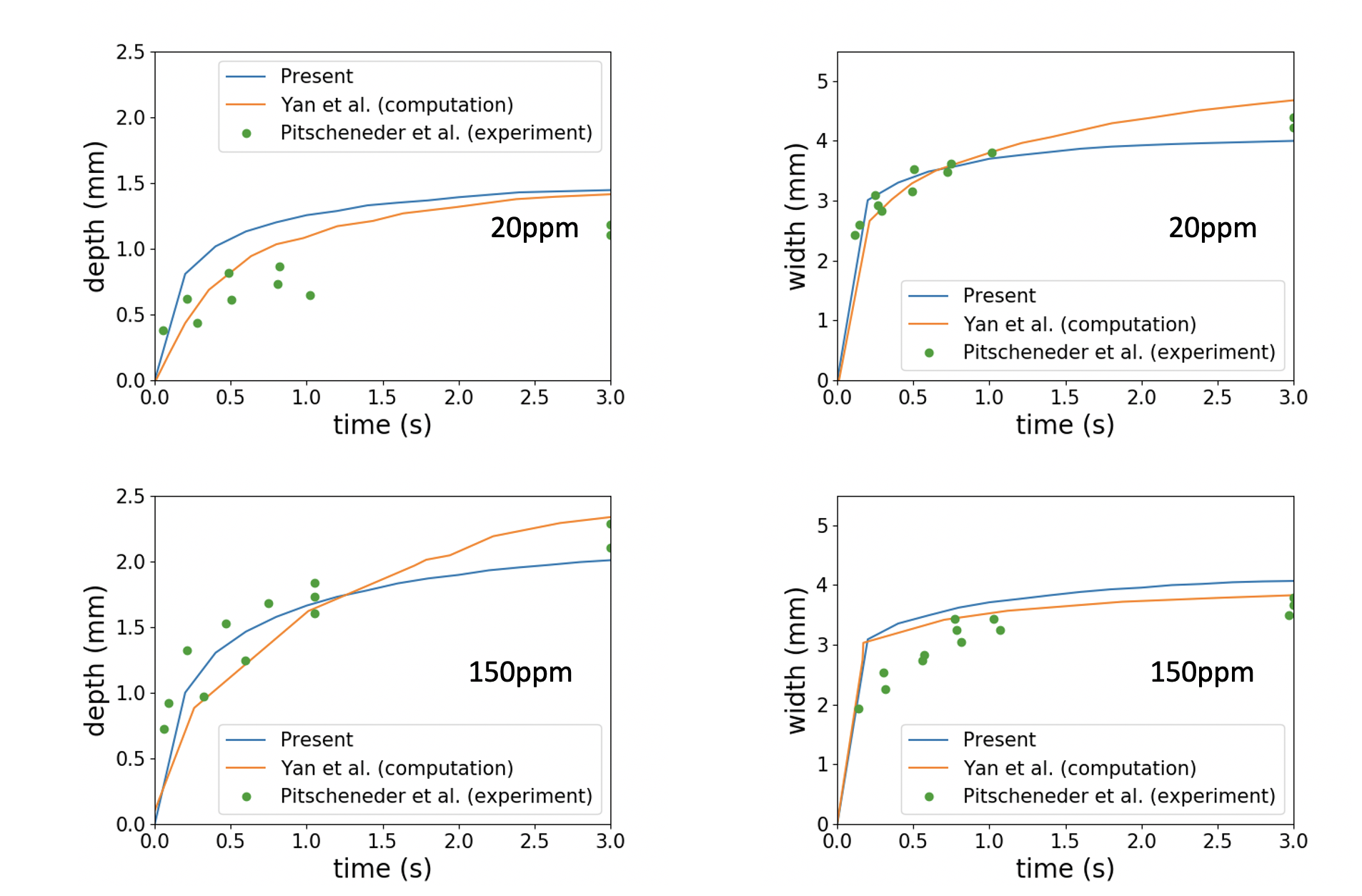}
 \caption{Time history of melt pool dimensions. Experimental results from~\cite{pitsch1996welding} and numerical predictions from~\cite{yan2018fully} are also plotted for comparison.}
 \label{fig:slm_hist}
\end{figure}

We simulate the problem with linear tetrahedral elements. As shown in Fig.~\ref{fig:slm_diagram} (right), the computational domain is a cylinder with a radius of $7\,\mathrm{mm}$ and a height of $20\,\mathrm{mm}$. The metal occupies the bottom half of the domain. A refined cylinder with a radius of $3\,\mathrm{mm}$ and a height of $12\,\mathrm{mm}$ is placed in the domain center to better capture the temperature and fluid dynamics. The mesh size gradually grows from the refined region to the outer boundaries from $0.08\,\mathrm{mm}$ to $2\,\mathrm{mm}$. \textcolor{black}{The total number of nodes and elements of the mesh are and 351,012 and 1,612,787, respectively.}

Since no material is deposited, a flat air-metal interface is assumed. The Marangoni effect, no penetration boundary condition, and heat source are applied on the air-metal interface by using the CSF models specified in Eq.~\ref{f_Marangoni} and Eq.~\ref{Q_laser_formula}. No heat flux and no-slip boundary conditions are applied on the three surfaces of the cylindrical domain. \textcolor{black}{The simulations were run with 192 processors with $\Delta t = 1\times 10^{-3}$ s}. This problem was investigated experimentally in~\cite{pitsch1996welding} and computationally in~\cite{yan2018fully} using a liquid-solid model. The results are used for comparison next.

Fig.~\ref{fig:slm_field1} and Fig.~\ref{fig:slm_field2} show the temperature contour, melt pool shape, and fluid velocity vectors scaled by their magnitude for the cases with two sulfur activities. The velocity vectors are tangential to the air-metal interface, which indicates that the no penetration boundary condition is enforced well by the CSF model. As shown in Fig.~\ref{fig:slm_mgn}, the Marangoni coefficient has different signs in the melt pools for $a_i$ = 20 ppm and $a_i$ = 150 ppm cases, which leads to different melt pool shape and opposite flow circulations, as depicted in Fig.~\ref{fig:slm_field1}. In the $a_i$ = 20 ppm case, the Marangoni coefficient is mainly positive in the melt pool. The higher surface force in higher temperature drives the flow from the boundary to the center and digs a narrow and deep melt pool (see Fig.~\ref{fig:slm_field2} (left)). In contrast, the Marangoni coefficient in the $a_i$ = 150 ppm case is mainly negative in the melt pool. The higher surface force in lower temperature drives the flow from the center to the boundary and results in a wide and shallow melt pool (see Fig.~\ref{fig:slm_field2} (right)). Fig.~\ref{fig:slm_hist} shows the time history of melt pool dimensions. Experimental measurements from~\cite{pitsch1996welding} and numerical predictions from~\cite{yan2018fully} are also plotted for comparison. Good agreements are obtained.
\section{Laser powder bed fusion: Validation with Argonne's experiments}\label{LPBF}
This section presents the applications to the developed methods to laser powder bed fusion processes validated by experimental data from Argonne National Lab using high-speed and high-resolution x-ray imaging.
\subsection{Static laser melting}\label{static}

\begin{table}[!t]
\caption{Mechanical properties of Ti-6Al-4V}
\label{tab:Ti-6Al-4V_mat}
\begin{tabular}{p{4cm}p{4cm}p{2cm}}
\hline\noalign{\smallskip}
Name & Notation (units) & Value  \\
\noalign{\smallskip}\svhline\noalign{\smallskip}
  Solid density & $\rho_s\;(kg\cdot m^{-3})$ & $4400$ \\
  Liquid density & $\rho_l\;(kg\cdot m^{-3})$ & $4400$ \\
  Gas density & $\rho_s\;(kg\cdot m^{-3})$ & $0.894$ \\
  Solidus temperature & $T_s\;(K)$ & $1878$   \\
  Liquid temperature & $T_l\;(K)$ & $1928$   \\
  Boiling temperature & $T_b\;(K)$ & $3533$ \\
  Solid specific heat capacity & $c_{ps}\;(J\cdot kg^{-1}\cdot K^{-1})$ & $670$   \\
  Liquid specific heat capacity & $c_{pl}\;(J\cdot kg^{-1}\cdot K^{-1})$ & $730$  \\
  Gas specific heat capacity & $c_{pg}\;(J\cdot kg^{-1}\cdot K^{-1})$ & $680$  \\
  Solid solid conductivity & $\kappa_{s}\;(W\cdot m^{-1}\cdot K^{-1})$ & $21$  \\
  Liquid solid conductivity & $\kappa_{l}\;(W\cdot m^{-1}\cdot K^{-1})$ & $29$   \\
  Gas solid conductivity & $\kappa_{g}\;(W\cdot m^{-1}\cdot K^{-1})$ & $0.028$   \\
  Latent heat of fusion & $L_m\;(kJ\cdot kg^{-1}\cdot K^{-1})$ & $290$   \\
  Latent heat of evaporation & $L_v\;(kJ\cdot kg^{-1}\cdot K^{-1})$ & $9600$   \\
  Solid viscosity & $\mu_s\;(Pa\cdot s)$ & $1.0\times 10^{6}$   \\
  Liquid viscosity & $\mu_l\;(Pa\cdot s)$ & $5.0\times 10^{-3}$   \\
  Gas viscosity & $\mu_l\;(Pa\cdot s)$ & $1.5\times 10^{-5}$   \\
  Surface tension coefficient & $\sigma_0\;(N\cdot m^{-1})$ & $1.68$ \\
  Stefan-Boltzmann constant & $\sigma_{SB}\;
  (kg \cdot s^{-3} \cdot K^{-4}) 
  $ & $5.67\times10^{-8}$\\
  Marangoni coefficient & $\frac{\partial \gamma}{\partial T}\;(N\cdot m^{-1} \cdot K^{-1})$ & $-2.6\times 10^{-4}$  \\
\noalign{\smallskip}\hline\noalign{\smallskip}
\end{tabular}
\end{table}

\begin{figure}[!htbp]
	\centering
	{\includegraphics[width=4.5in]{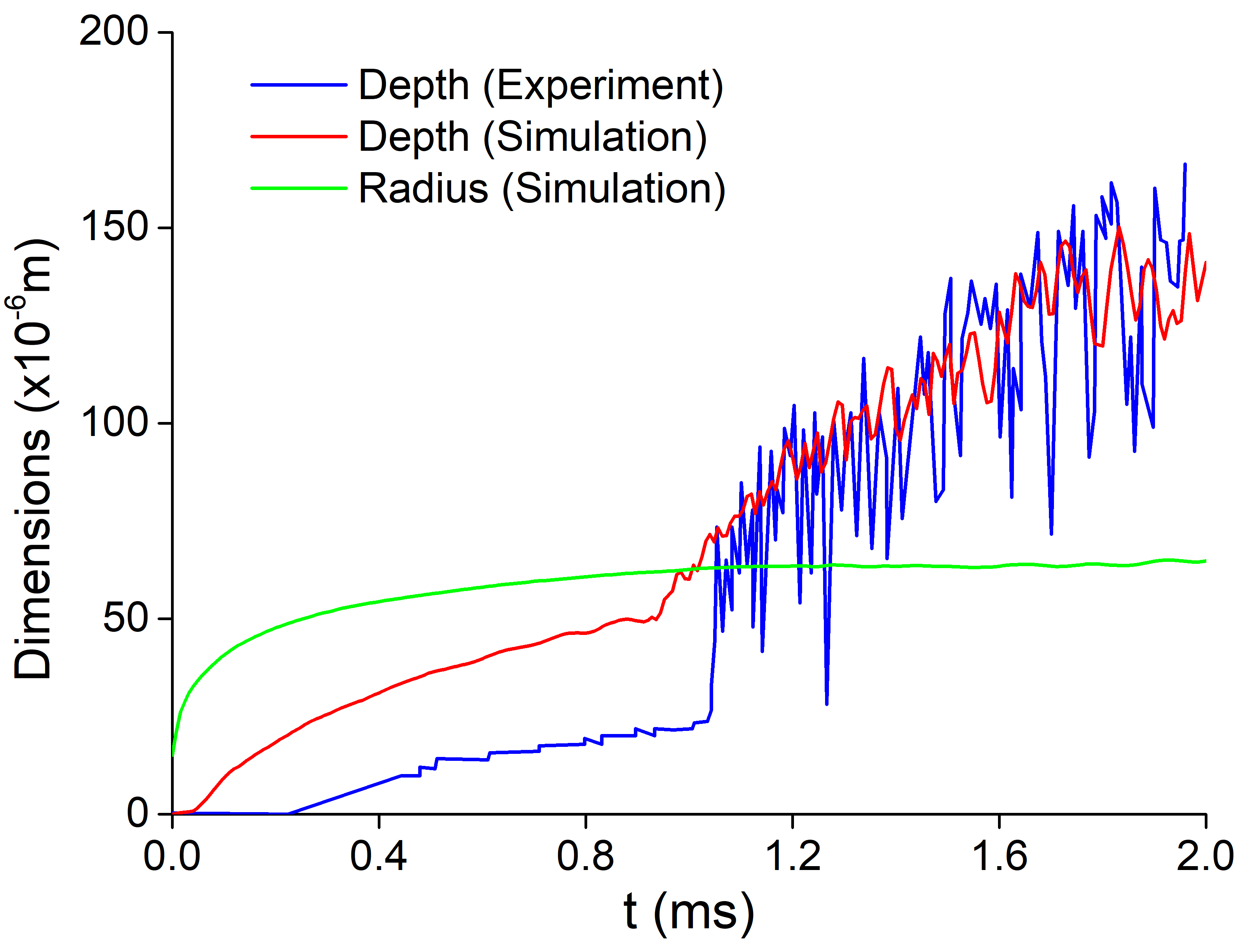}}
	\caption{Time history of melt pool depth in the stationary case.}
	\label{fig:case3_depth_static}
\end{figure}

\begin{figure}[!htbp]
	\centering
	{\includegraphics[width=4.5in]{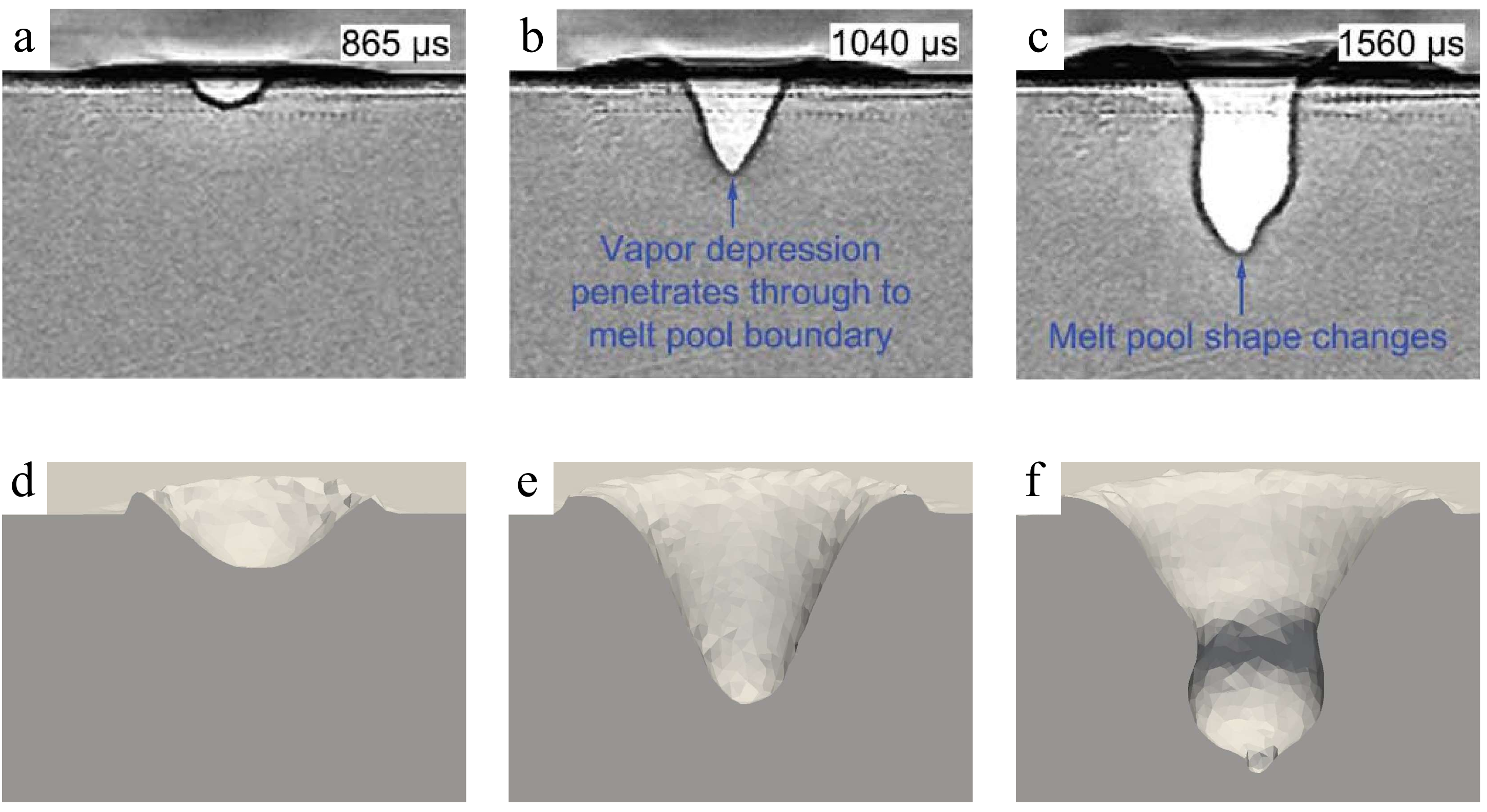}}
	\caption{Comparison of keyhole shapes in the stationary case between experimental images and current simulation. Top: experiment. Bottom: current simulation.}
	\label{fig:case3_phi_comparison_static}
\end{figure}

\begin{figure}[!htbp]
	\centering
	{\includegraphics[width=4.5in]{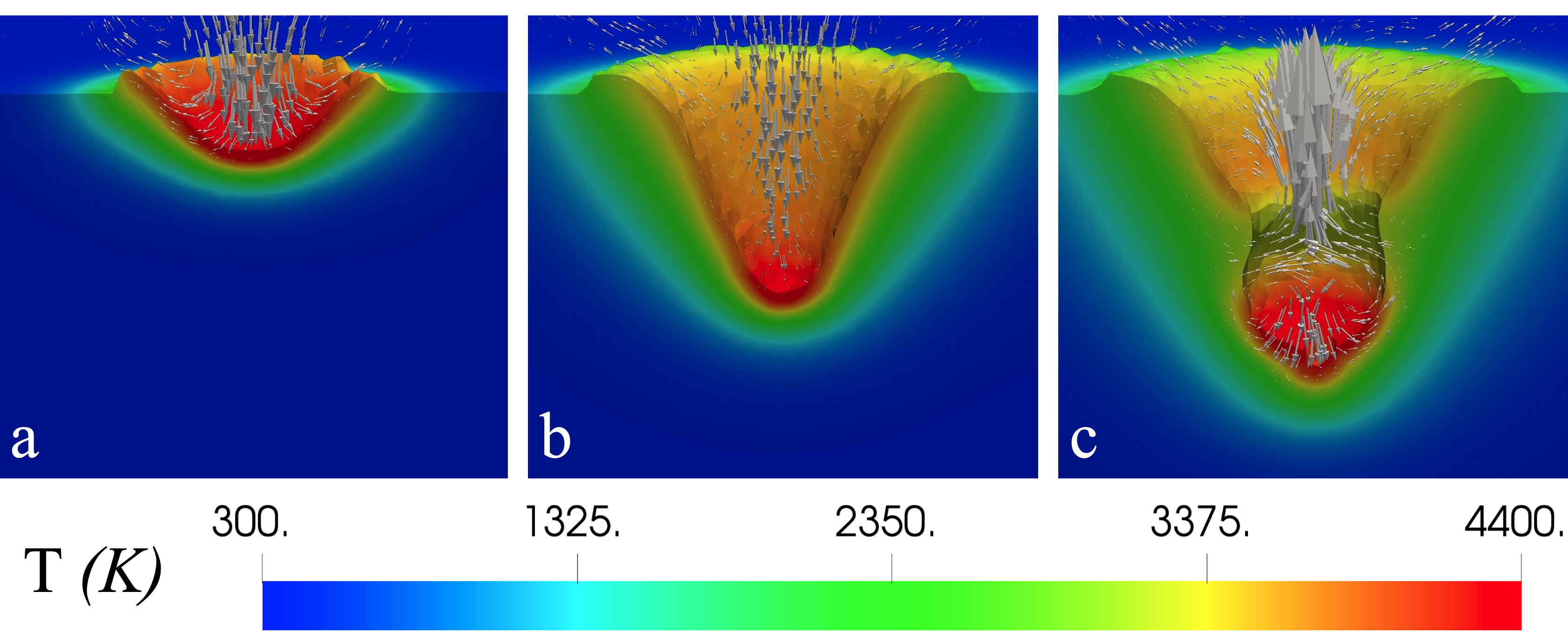}}
	\caption{Temperature in the metal phase and gas velocity vectors.}
	\label{fig:case3_tem_static}
\end{figure}
\begin{figure}[!htbp]
	\centering
	{\includegraphics[width=4.5in]{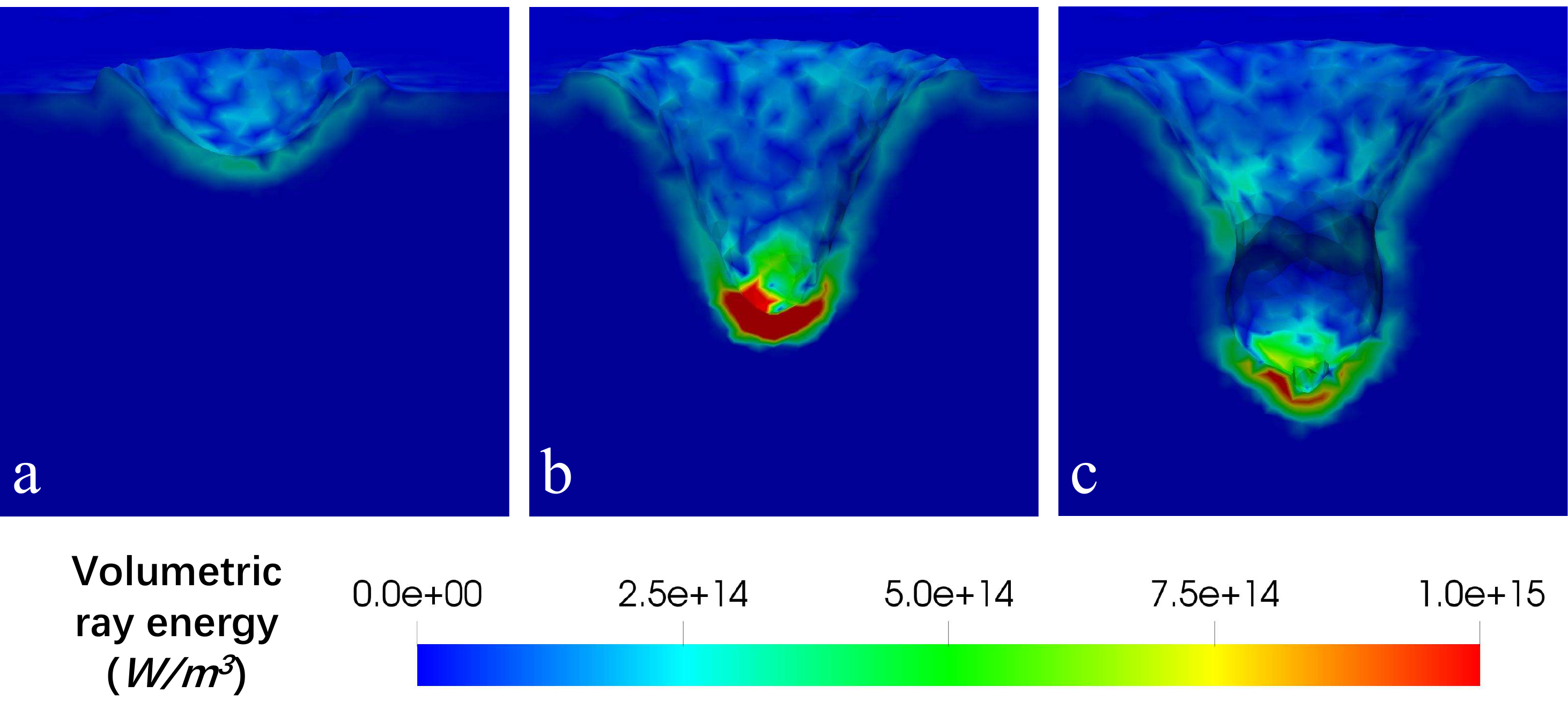}}
	\caption{Volumetric ray energy on the melt pool surface.}
	\label{fig:case3_laser_energy_static}
\end{figure}

\begin{figure}[!htbp]
	\centering
	{\includegraphics[width=4.5in]{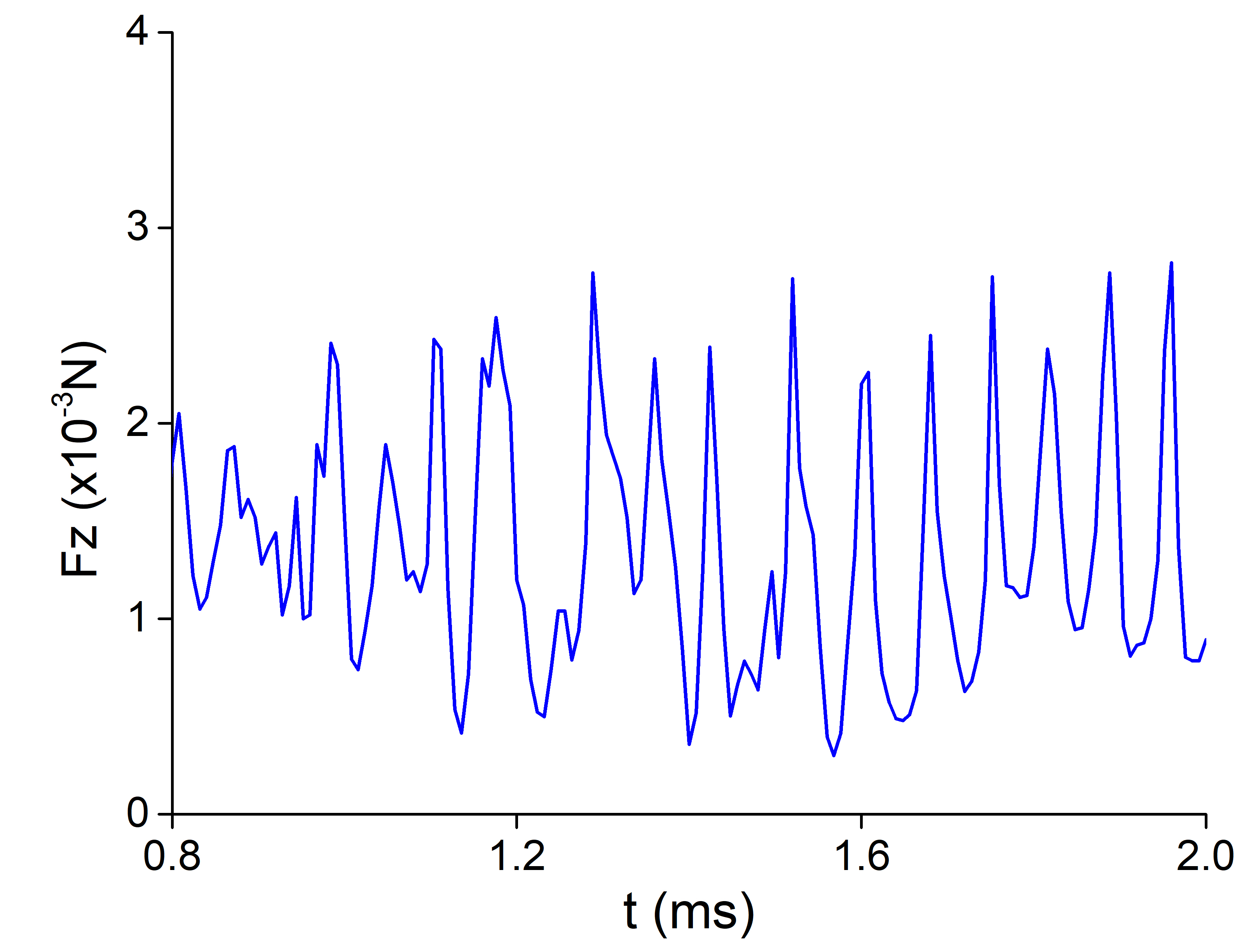}}
	\caption{Time history of recoil force $F_z$ on the melt pool surface.}
	\label{fig:case3_fz_static}
\end{figure}

\begin{figure}[!htbp]
	\centering
	{\includegraphics[width=4.5in]{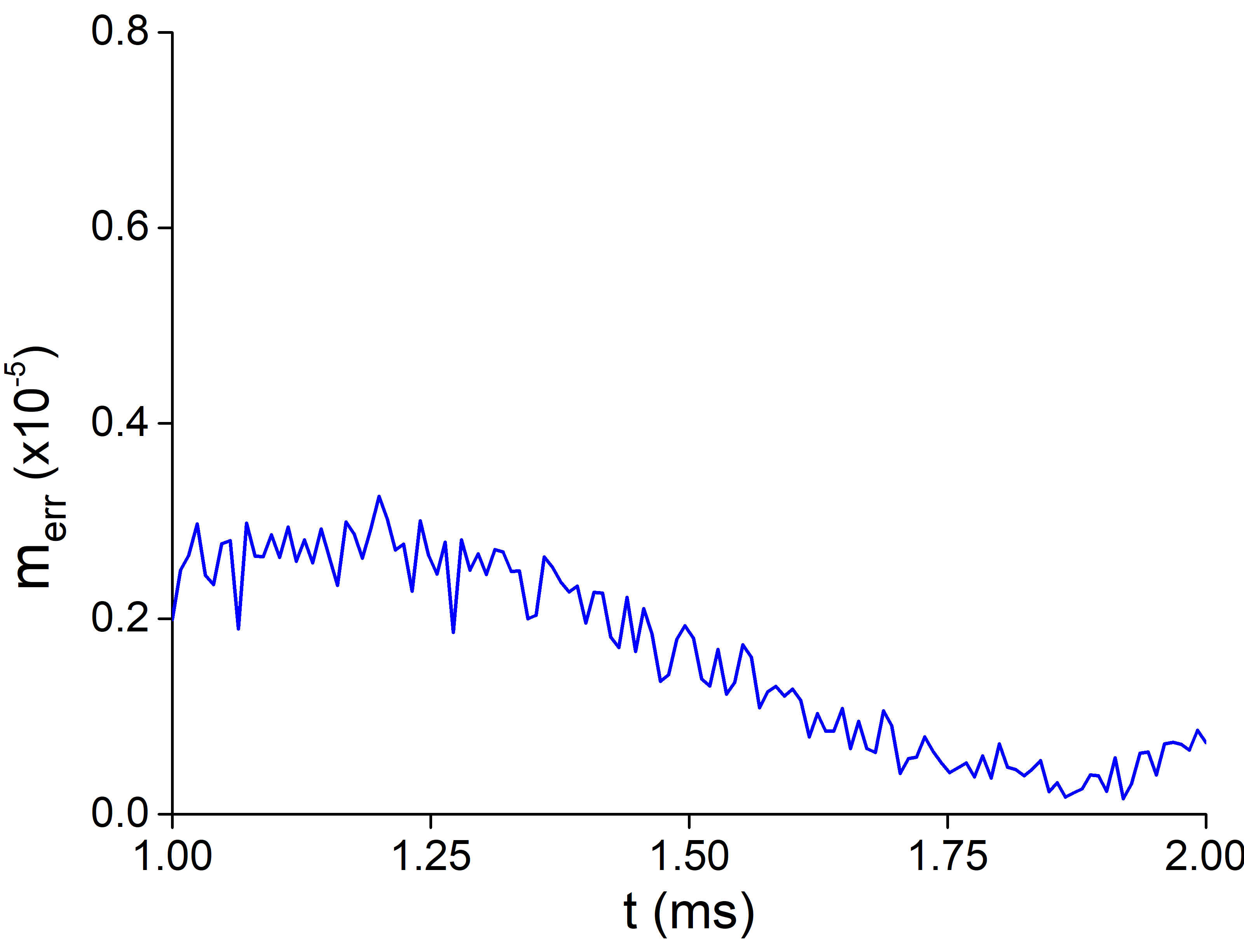}}
	\caption{Time history of relative mass conservation error.}
	\label{fig:case3_redist_loss_static}
\end{figure}

For metal AM applications, we first simulate a stationary laser melting problem to demonstrate the modeling capabilities of phase transition and keyhole evolution. In the simulation, Argon and Ti-6Al-4V are used for the gas and metal phases, respectively, and their mechanical and thermal properties, extracted from~\cite{klassen2014evaporation,yan2017multi,bayat2019keyhole,tan2013investigation,wang2020evaporation}, are listed in Table~\ref{tab:Ti-6Al-4V_mat}. The laser spot size of laser power are 140 $\mu$m and $Q =$ 156 $W$. The simulation is performed on a cubic domain with unstructured tetrahedral elements. A refined region with element length $h$ = 3.0 $\mu m$ around the melt pool is designed to better capture the dynamics. The total number of elements and nodes are 2,466,919 and 428,566, respectively. No-penetration and fixed temperature boundary conditions are adopted for all the surfaces. The simulation runs for 2 $ms$ with $\Delta t = 5.0 \times 10^{-7}$ $s$.

This problem was experimentally investigated by Argonne National Laboratory using ultrahigh-speed x-ray imaging~\cite{cunningham2019keyhole}. Instead of focusing on physics discussions, here we focus on using the available quantitative experimental results to validate the simulated results and report the quantities that experiments cannot measure. Fig.~\ref{fig:case3_depth_static} shows the time history of melt pool radius and penetrating depth. The melt pool radius is measured by the averaged distance from the intersection between solid-liquid/gas-metal interfaces to the laser center, while the penetrating depth is measured by the distance from the deepest point of the melt pool to the still gas-metal interface. The predicted penetrating depth shows a reasonable agreement with the experimental measurements, especially for the drilling rate in the keyhole instability stage. The simulation also generates similar keyhole shapes to the experimental images on the middle plane, as seen in Fig.~\ref{fig:case3_phi_comparison_static}. However, the fluctuation of penetrating depth predicted by the simulation is smaller than the experimental results, which may be related to the empirical parameters in the evaporation model.

The following quantities that experiments cannot provide are reported. Fig.~\ref{fig:case3_tem_static} depicts the temperature distribution in the metal phase and velocity vectors in the gas phase at three instances, and the corresponding volumetric distribution of ray intensities (multiplied with density scaled delta function $\delta_s$) is presented in Fig.~\ref{fig:case3_laser_energy_static}. The recoil force $F_z$ is one critical factor controlling keyhole dynamics in metal AM. In the context of interface capturing, $F_z$ is computed as 

\begin{align}
\label{Fz_for}
F_z = -\int_{\Gamma_I} P_{recoi} d\Gamma = -\int_{\Omega} P_{recoi} \delta_s d\Omega
\end{align}
The time history of $F_z$ in the keyhole instability stage is showed in Fig.~\ref{fig:case3_fz_static}. The magnitude and fluctuation range agrees with what was reported in~\cite{wang2020evaporation}. Mass conversation is critical in multi-phase flows simulations. In this work, the relative global metal mass error $m_{err}$ is quantified as 

\begin{align}
\label{mass_err_formula}
m_{err} = \frac{m_0 - m_t - \int_0^T \int_{\Gamma_I} m_{e} d\Gamma dt}{m_0}
\end{align}
where $m_0$ is the initial metal mass, $m_t$ is the metal mass at current time, evaluated as $m_t = \int_\Omega \rho_m H(\phi) d\Omega $, the last term in the numerator is the accumulated evaporated metal mass up to current time. The time history of $m_{err}$ is plotted in Fig.~\ref{fig:case3_redist_loss_static}, which shows the relative metal mass error is maintained at an order of $10^{-6}$.
\begin{figure}[!htbp]
	\centering
	{\includegraphics[width=4.5in]{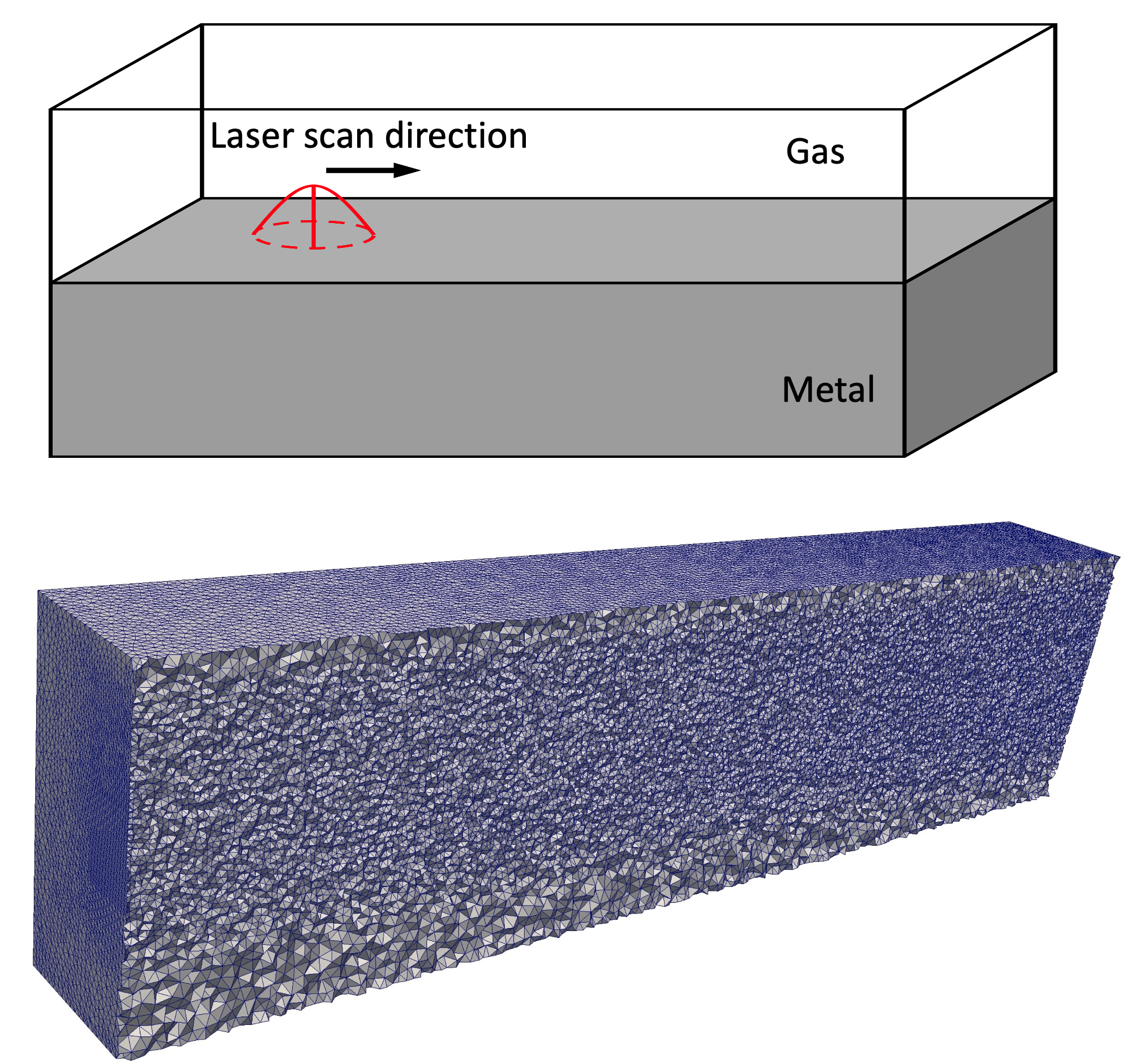}}
	\caption{Simulation setup and mesh employed in the moving laser case.}
	\label{fig:case4_laser_energy_moving}
\end{figure}

\begin{figure}[!htbp]
	\centering
	{\includegraphics[width=4.5in]{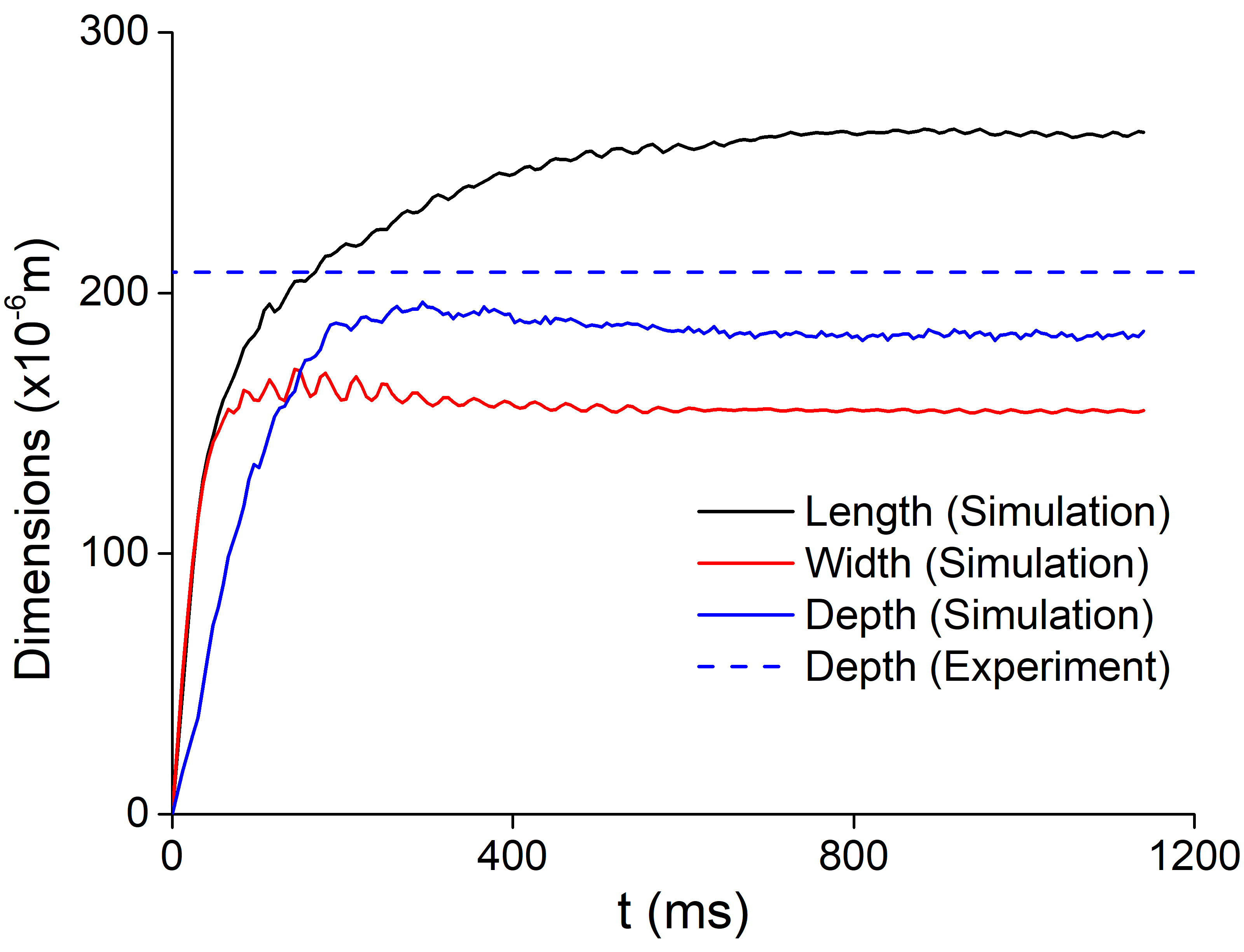}}
	\caption{Time history of melt pool dimensions in the moving laser case. The time averaged melt pool depth from~\cite{cunningham2019keyhole} is plotted for comparison. The relative discrepancy in terms of depth between simulation and experiment is 10.3 $\%$.}
	\label{fig:case4_depth_moving}
\end{figure}

\begin{figure}[!htbp]
	\centering
	{\includegraphics[width=4.5in]{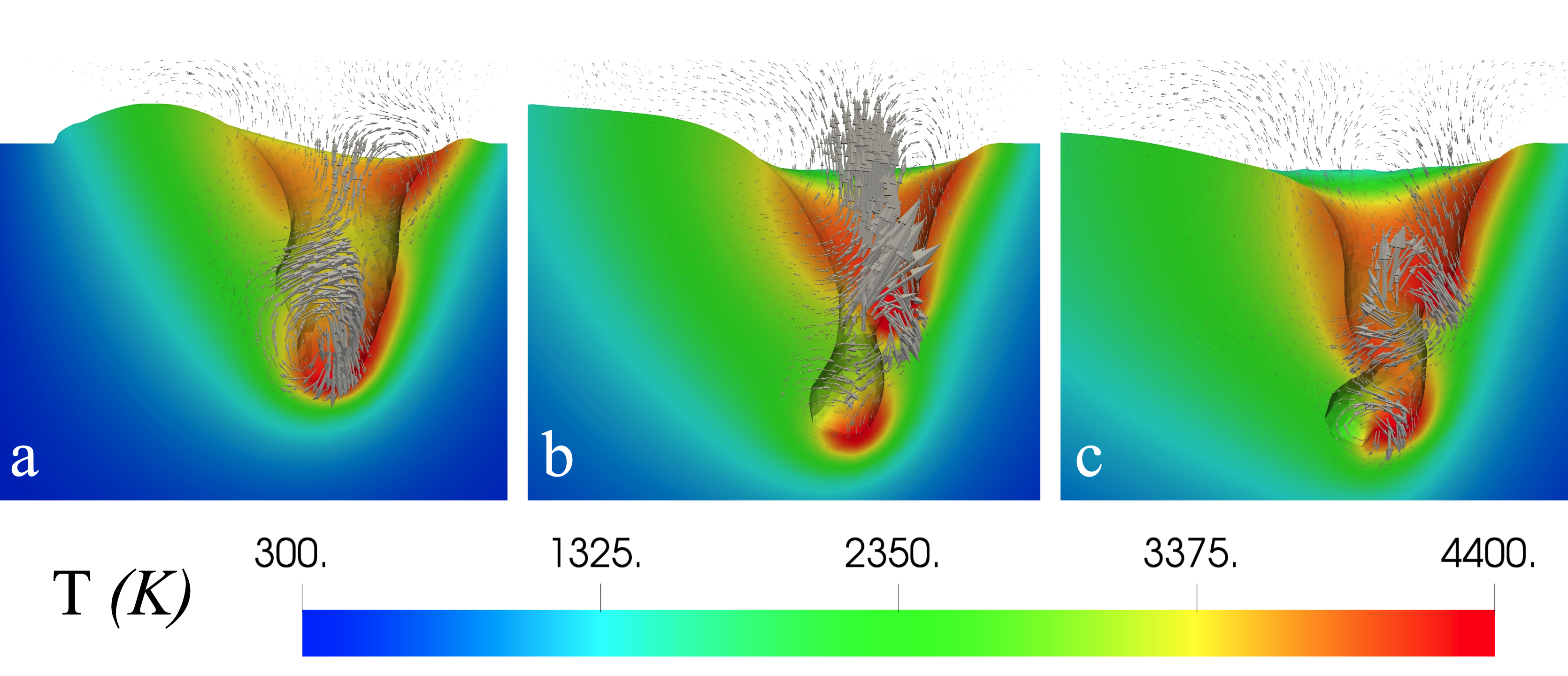}}
	\caption{Melt pool shapes, temperature in the metal, and gas velocity vectors in the moving laser case.}
	\label{fig:case4_tem_moving}
\end{figure}

\begin{figure}[!htbp]
	\centering
	{\includegraphics[width=4.5in]{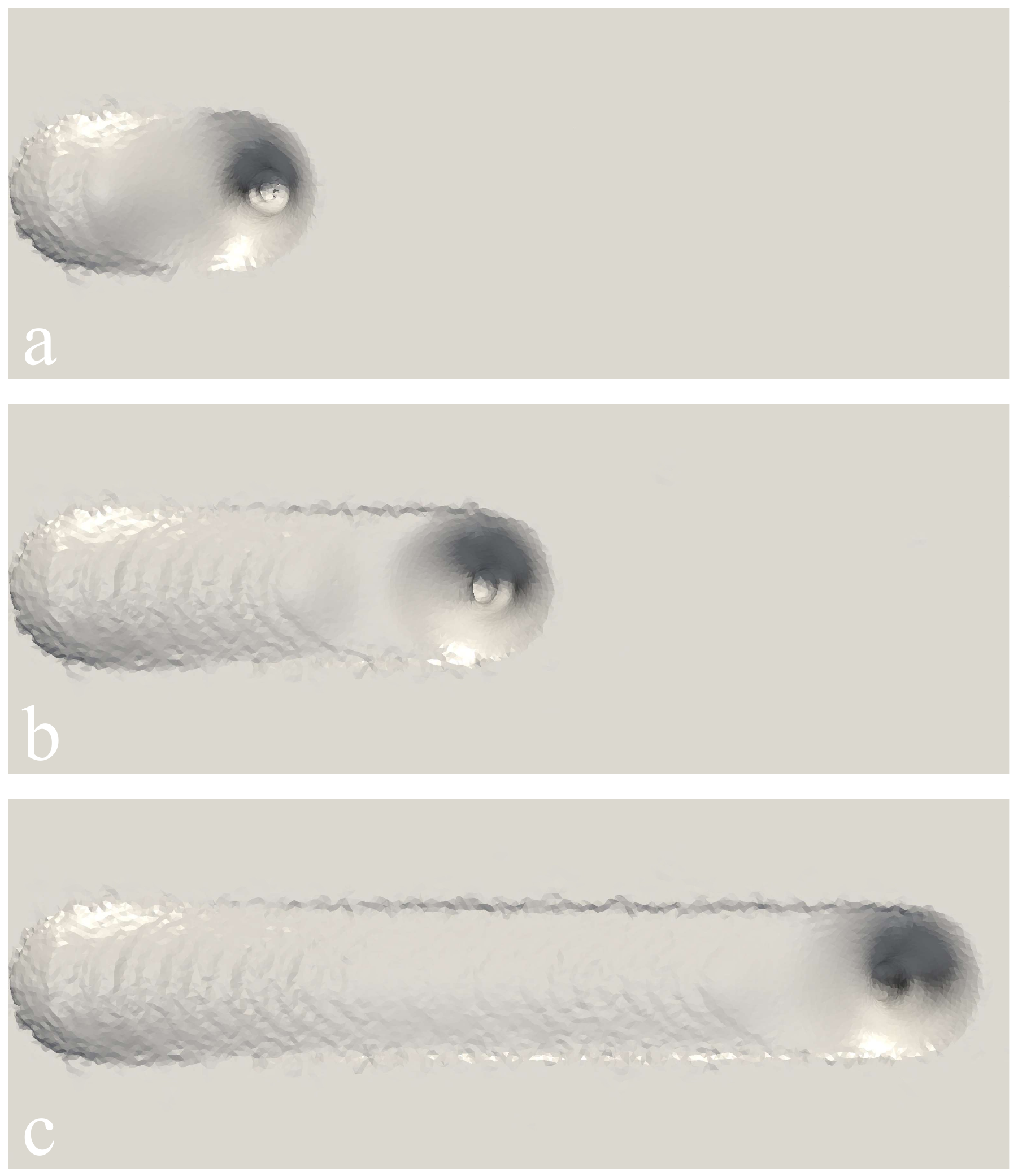}}
	\caption{Topography of metal top surface in the moving laser case.}
	\label{fig:case4_top_view_moving}
\end{figure}

\begin{figure}[!htbp]
	\centering
	{\includegraphics[width=4.5in]{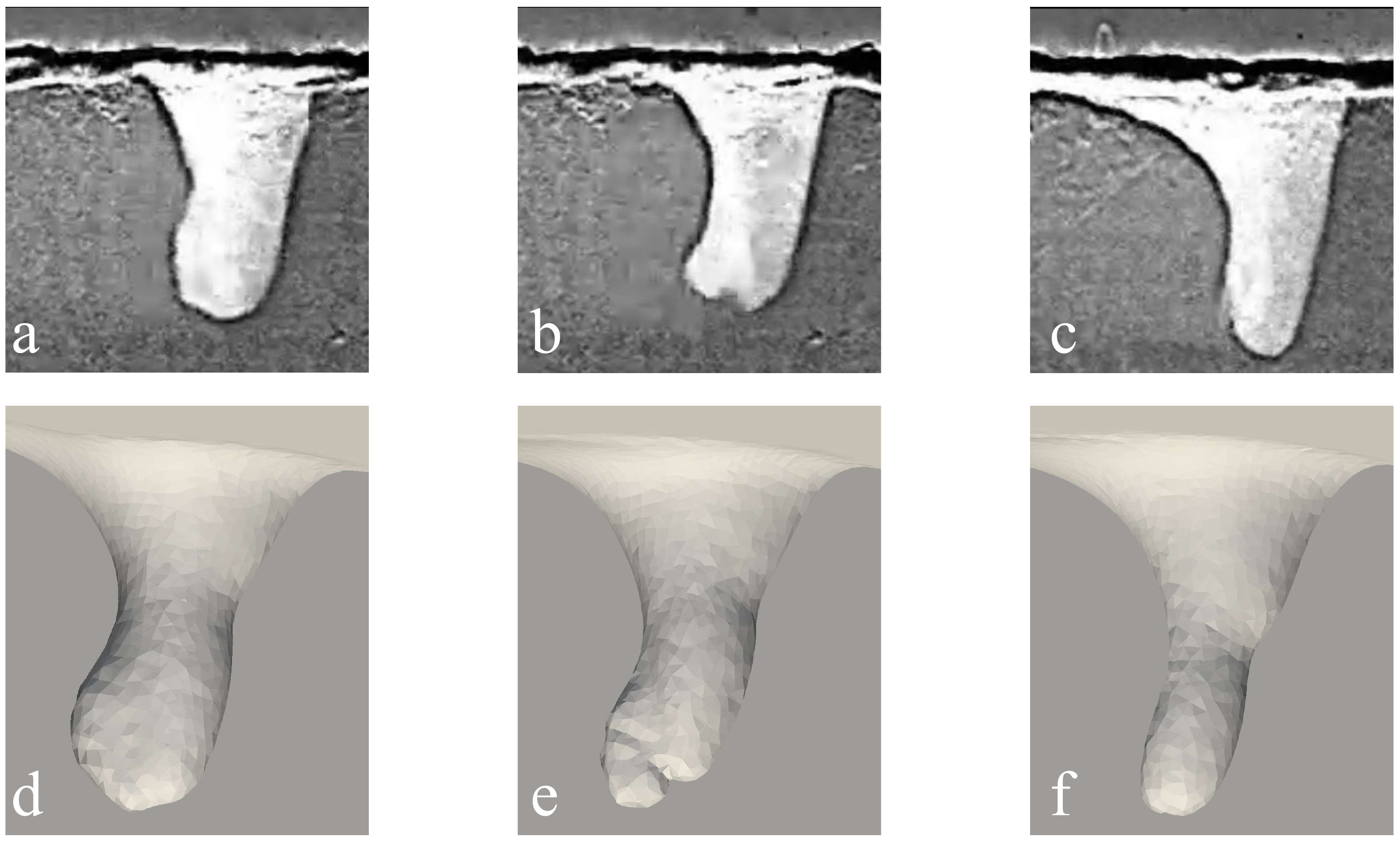}}
	\caption{Comparison of keyhole shapes between experimental images and current simulation in the moving laser case. The average front keyhole wall angle predicted from the simulation is $69.8^{\circ}$, compared with $76.0^{\circ}$ from the experiment in~\cite{cunningham2019keyhole}.}
	\label{fig:case4_phi_comparison_moving}
\end{figure}

\begin{figure}[!htbp]
	\centering
	{\includegraphics[width=4.5in]{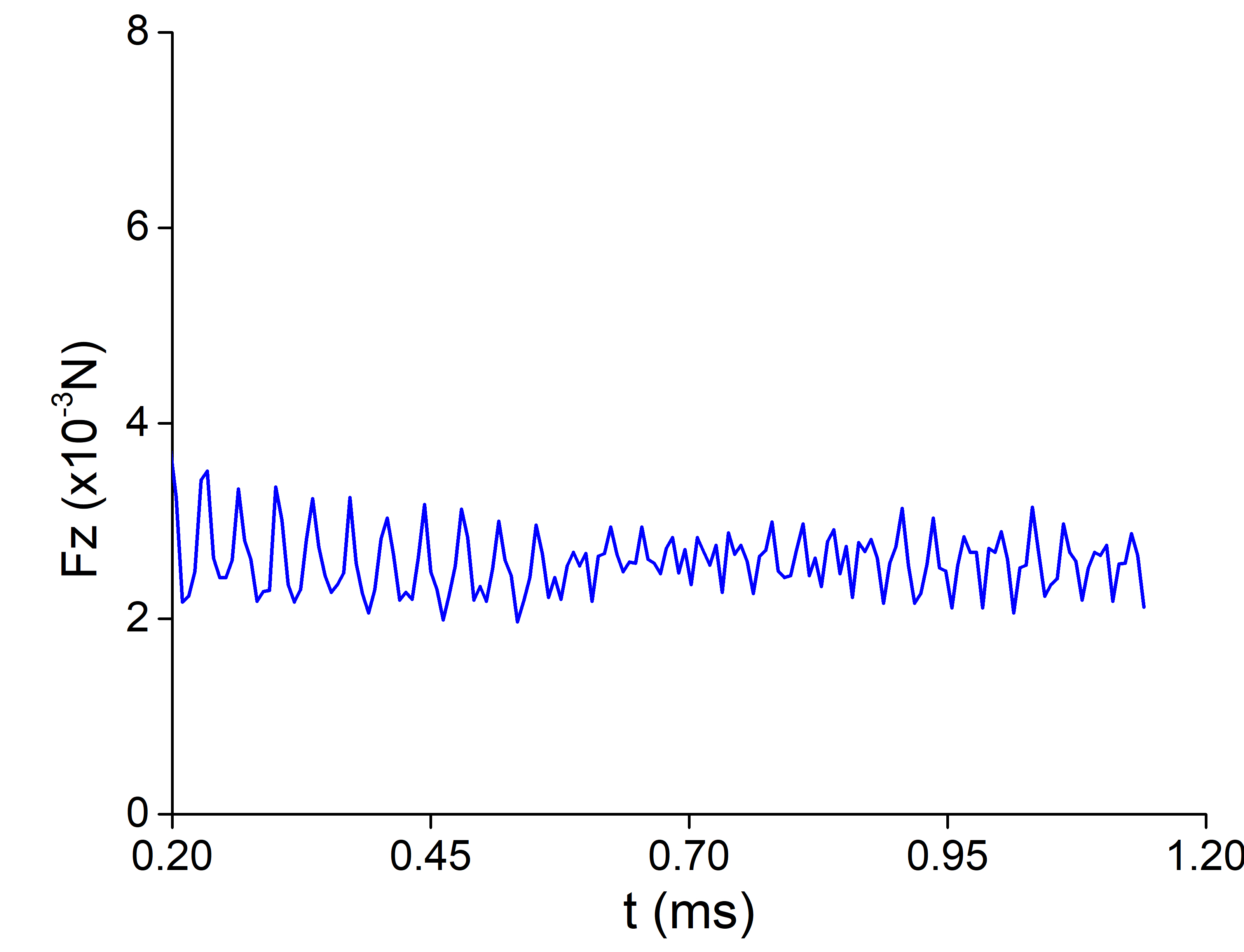}}
	\caption{Time history of recoil force $F_z$ in the moving laser case.}
	\label{fig:case4_fz_moving}
\end{figure}

\begin{figure}[!htbp]
	\centering
	{\includegraphics[width=4.5in]{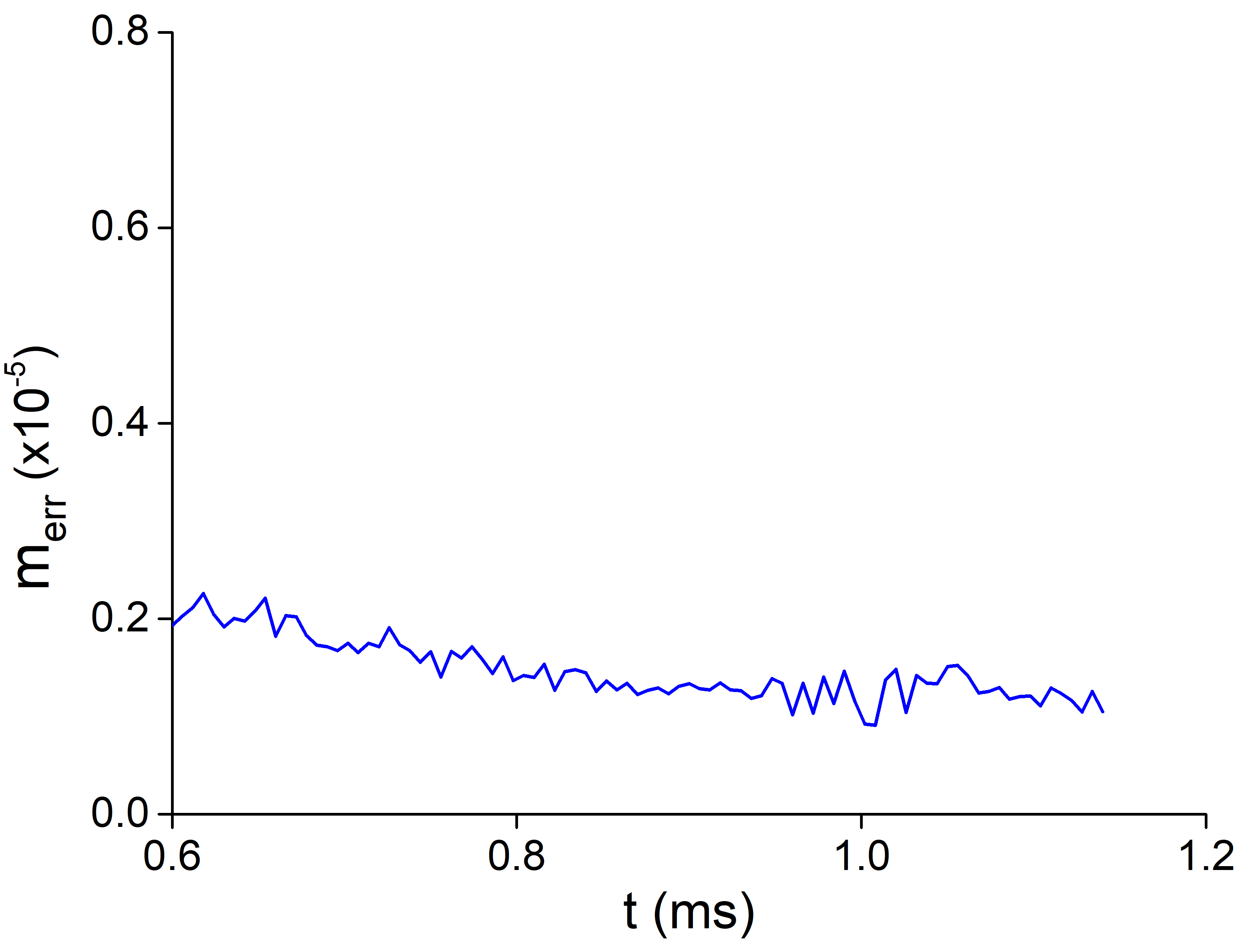}}
	\caption{Relative metal mass conservation error in the moving laser case.}
	\label{fig:case4_redist_loss_moving}
\end{figure}

\subsection{Moving laser melting}\label{moving}

We further test the proposed methods by simulating a moving laser case with a Ti-6Al-4V bare plate, which is also experimentally investigated in~\cite{cunningham2019keyhole}. The laser spot size is 95 $\mu m$, the laser power is $Q$ = 364 $W$, and the scan speed is $V_s$ = 900 $mm/s$. The computation is performed in a box with a refined region around the laser track with element length $h$ = 3 $\mu m$. The mesh has 4,215,023 elements and 612,754 nodes in total. $\Delta t = 2.5 \times 10^{-7}$ $s$. Fig.~\ref{fig:case4_laser_energy_moving} shows the problem setup and the mesh employed in the simulation. The same types boundary condition as the stationary laser case are used. 

Fig.~\ref{fig:case4_depth_moving} shows the time history of the melt pool dimensions. The time-averaged experimental melt pool depth~\cite{cunningham2019keyhole} is also plotted for comparison. The relative discrepancy in depth between simulation and experiment is less than 10.3 $\%$. Compared with the stationary laser case, the depth fluctuation is smaller. This agrees with the trend in Argonne National Lab's latest high-speed imaging experiments~\cite{zhao2020critical}, which found the relative fluctuation of melt pool depth decreases with increasing laser scan speed. Although the depth fluctuation is small, the keyhole instability is still pronounced, resulting in violent free surface deformation, as seen in Fig.~\ref{fig:case4_tem_moving}, which shows the melt pool shape, temperature field in the metal, and gas velocity vectors. From Fig.~\ref{fig:case4_tem_moving}, we also observe that the heat-induced gas velocity is more turbulent compared with the stationary laser case because of the more significant variation of interfacial forces induced by the moving laser. In particular, as we show in Fig.~\ref{fig:case4_top_view_moving}, the simulation captures the common experimentally observed chevron-type topography, primarily induced by Marangoni force, on the metal top surface.  Fig.~\ref{fig:case4_phi_comparison_moving} shows the comparison of keyhole shapes between experimental images and the current simulated results at three time instances. Similar shapes are obtained. The averaged front keyhole wall angle predicted from the simulation is $69.8^{\circ}$, compared with $76.0^{\circ}$ reported from the experiment~\cite{cunningham2019keyhole}.

Fig.~\ref{fig:case4_fz_moving} shows the time history of recoil force integrated over the melt pool surface. The averaged magnitude of $F_z$ is in the same order as that of the stationary case, but with smaller fluctuation, which qualitatively explains the smaller depth fluctuation seen in Fig.~\ref{fig:case4_depth_moving}. The time history of relative mass conservation error is in the same order as the stationary laser case, as shown in Fig.~\ref{fig:case4_redist_loss_moving}.
\newpage
\section{Directed energy deposition (DED)}\label{DED}

\subsection{Deposit geometry}\label{deposit}
\begin{figure}[!htbp]
 \centering
 \includegraphics[scale=0.3]{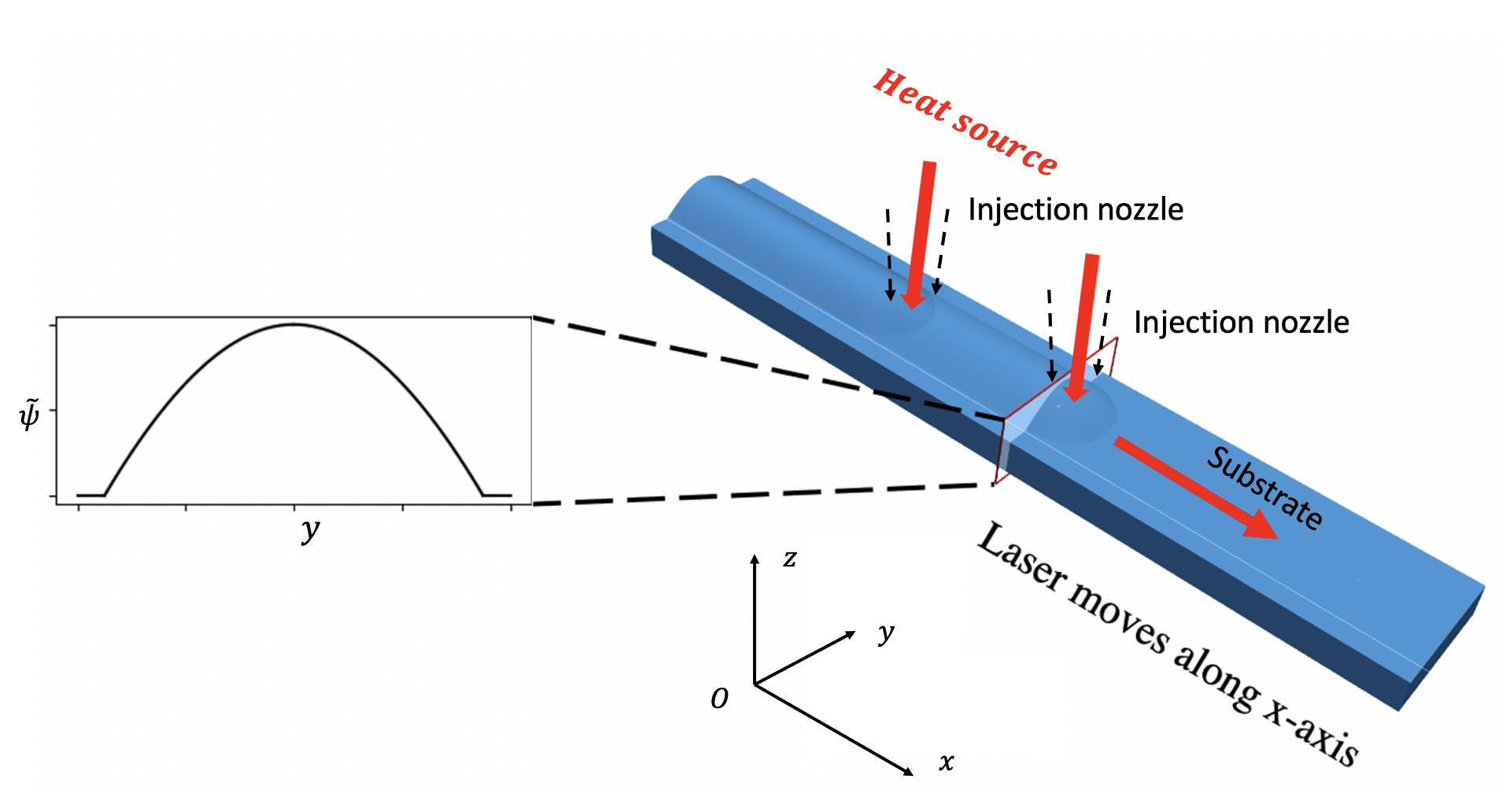}
 \caption{Diagram of a single-track DED process. As the laser moves, the surrounding nozzles simultaneously inject metallic powders into the laser beam and deposit them onto the substrate.}
 \label{fig:ded_process}
\end{figure}

Deposit geometry is an important factor for DED simulations. Previous models either rely on experimental observations or presumed 2D/3D geometry shapes, such as parabolic, sinusoidal or elliptical surfaces~\cite{cao2011curve,corbin2017effect,xiong2014curve,doumanidis2002multivariable,debroy2017digital}. These models are simple to implement but lack a sound physical foundation. Departing from these models, we propose a new theoretical approach to derive the deposit geometry based on an energy minimization problem with mass conservation constraint. Fig.~\ref{fig:ded_process} shows the schematic diagram of a typical single-track DED process. The proposed approach makes two assumptions: (1) The cross-section of the deposit remains unchanged behind the laser. (2) The material distribution is radially symmetric under the laser beam. To derive the deposit shape, we first define the following energy function in the deposit volume.
\begin{align}\label{e_phi}
 E &=\int_{\mathbf{\Gamma}_t}\sigma d\Gamma + \int_{\mathbf{\Omega}_t^{ded}} \rho_m g z d {\Omega}
\end{align}
where $\mathbf{\Omega}_t^{ded}$ is the volume occupied by the material deposit, $\mathbf{\Gamma}_t$ is the surface of $\mathbf{\Omega}_t^{ded}$ exposed to the air. In Eq.~\ref{e_phi}, the first term represents the surface energy, where $\sigma$ is the surface tension coefficient. The second term is the gravitational potential energy, where $\rho_m$ is the metal density, $g$ is the gravitational acceleration magnitude, $z$ is the height with respect to the substrate. The deposit geometry $\mathbf{\Gamma}_t$ can be described as a height function $\psi(x, y)$ with respect to the substrate ($x$-$y$ plane), namely,
\begin{align}
\mathbf{\Gamma}_t=\{\mathbf{x}=(x, y, z)^T|z=\psi(x, y), \mathbf{x}\in\mathbb{R}^3\}
\end{align}
Then, the total energy function in Eq.~\eqref{e_phi} can be expressed as
\begin{align}\label{e_phi1}
    E (\psi) =\int_{\mathbf{\Gamma}_t^p}(\sigma \sqrt{1+\psi,_{x}^2+\psi,_{y}^2}  + \frac{1}{2}\rho_m g \psi^2 )dxdy
\end{align}
where $\mathbf{\Gamma}_t^p$ is the projection of $\mathbf{\Gamma}_t$ on the substrate. $\psi(x, y)$ can be obtained by minimizing the energy function $E (\psi)$ subject to appropriate constraints. Considering the caught mass from the nozzle must equal the mass in the deposit volume, the following constraint needs to be satisfied, $\rho_m V_s A =\eta_c \dot{m}$, where $A$ is the area of the cross-section behind heat laser, $\eta_c$ is the fractional mass catchment of material into the melt pool, $\dot{m}$ is the mass flow rate from the nozzle, $V_s$ is the scanning speed of the laser. Considering the facts that the deposit cross-section behind the laser remains unchanged and the deposit length is much longer than its width, it is reasonable to assume ${\psi,_{x}}=0$ in Eq.~\ref{e_phi1}, which reduces the 3D minimization problem to the following 2D minimization problem
\begin{align}\label{eq:minimization_2D}
& \tilde{\psi} = \underset{\tilde{\psi}}{\text{argmin}}
& &\tilde{E}(\tilde{\psi}) = \int^{\frac{L}{2}}_{-\frac{L}{2}}(\sigma \sqrt{1+\tilde{\psi},_{y}^2} + \frac{1}{2}\rho_m g\tilde{\psi}^2) dy \\
& \text{subject to}
& &  \nonumber \rho_m V_s \int^{\frac{L}{2}}_{-\frac{L}{2}}\tilde{\psi} dy=\eta_c \dot{m}
\end{align}
 where $\tilde{\psi} = \tilde{\psi} (y)$ is the height function of the deposit cross-section behind the laser (see Fig.~\ref{fig:ded_process}), $L$ is the deposit width, which is a fraction $f_m$ of the laser beam radius $r_b$. $f_m$ varies from 0.75 and 1, depending on the manufacturing parameters~\cite{debroy2017digital}. 
 
 The minimization problem in Eq.~\ref{eq:minimization_2D} can be solved by using a Lagrangian multiplier approach, in which the following Lagrangian functional is defined
\begin{align}
 F (\tilde{\psi}) &= \tilde{E}(\tilde{\psi}) + \lambda \left(\rho_m V_s \int^{\frac{L}{2}}_{-\frac{L}{2}}\tilde{\psi} dy-\eta_c \dot{m}\right)
\end{align}
where $\lambda$ is an unknown Lagrangian multiplier. The stationary point is obtained by setting $\frac{\delta F}{\delta \tilde{\psi}} = 0$ and $\frac{\delta F}{\delta \lambda} = 0$, which leads to the following two Euler-Lagrange equations with boundary conditions. 
\begin{align}
&\tilde{\psi},_{yy}=\frac{\frac{\rho g\tilde{\psi}+\lambda}{\sigma}\sqrt{1+\tilde{\psi},_y^2} + \tilde{\psi},_y^2}{1+\tilde{\psi},_y^2}\label{phi_geo}\\
&\int^{\frac{L}{2}}_{-\frac{L}{2}}\tilde{\psi} dy = \frac{\eta_c \dot{m}}{\rho_m V_s} \label{phi_lambda}\\
&\tilde{\psi}(\pm\frac{L}{2}) = 0
\end{align}
This is a highly nonlinear ordinary differential equation (ODE) and has to be solved numerically. The good news is that we only need to solve it once. Once $\tilde{\psi}$ is given, the deposit geometry $\mathbf{\Gamma}_t$ is constructed as follows. The deposit behind the laser is obtained by extruding $\tilde{\psi}$ along $x$ direction, and the deposit front is obtained by rotating $\tilde{\psi}$ around the laser. With $\mathbf{\Gamma}_t$, a signed distance function that moves with the laser at the same speed is constructed to represent the air-metal interface. Then the approach and the associated methods of enforcement boundary conditions on the air-metal interface can be applied using the methods described in Section~\ref{sec: computation}.

\subsection{Direct energy deposition of SS-316L}
A single-track direct energy deposition (DED) process is simulated to demonstrate the proposed approach's predictive capability. The problem is set up as follows. A laser with a Gaussian profile scans across a flat SS-316L substrate with an initial temperature of $T_0=300$ $\textcolor{black}{\mathrm{K}}$. During the scanning, the nozzle around the laser simultaneously release SS-316L powders into the laser beam and deposit them onto the substrate. When the particles reach the built surface, it is assumed that they have been heated up to the local temperature. Thus, the absorbed energy by the depositing material can be computed as
\begin{align}
 & Q_v=\dot{m}\int_{T_0}^T c_{p}(\tau)d\tau
 \end{align}
The remaining laser energy entering the metal through the CSF model is
\begin{align}
 &  Q_T =\frac{d(\eta_p Q - Q_v)}{{\pi}r_b^2}exp(-d\frac{||\mathbf{x}-\mathbf{x_c}||^2}{r_b^2}) (\mathbf{n}\cdot\mathbf{e}_3) \delta_\epsilon
\end{align}

The properties of SS-316L and manufacturing parameters utilized in the paper are listed in Table~\ref{tab:ded_param} and Table~\ref{tab:dedproc_param}. The simulations make use of a box with dimensions of $30.0 \times 5.0 \times 4.2$ mm. \textcolor{black}{Structured hexahedral elements are used.} A refined region with element length $0.06$ mm is designed along the track to better capture the temperature and fluid dynamics. \textcolor{black}{The generated mesh consists of 496,571 nodes and 388,960 elements.} The simulation is performed with $\Delta t $ = $0.25\times 10^{-3}$ s until the melt pool reaches the quasi-steady state.

\begin{figure}[!htbp]
   \centering
   \includegraphics[width=\linewidth]{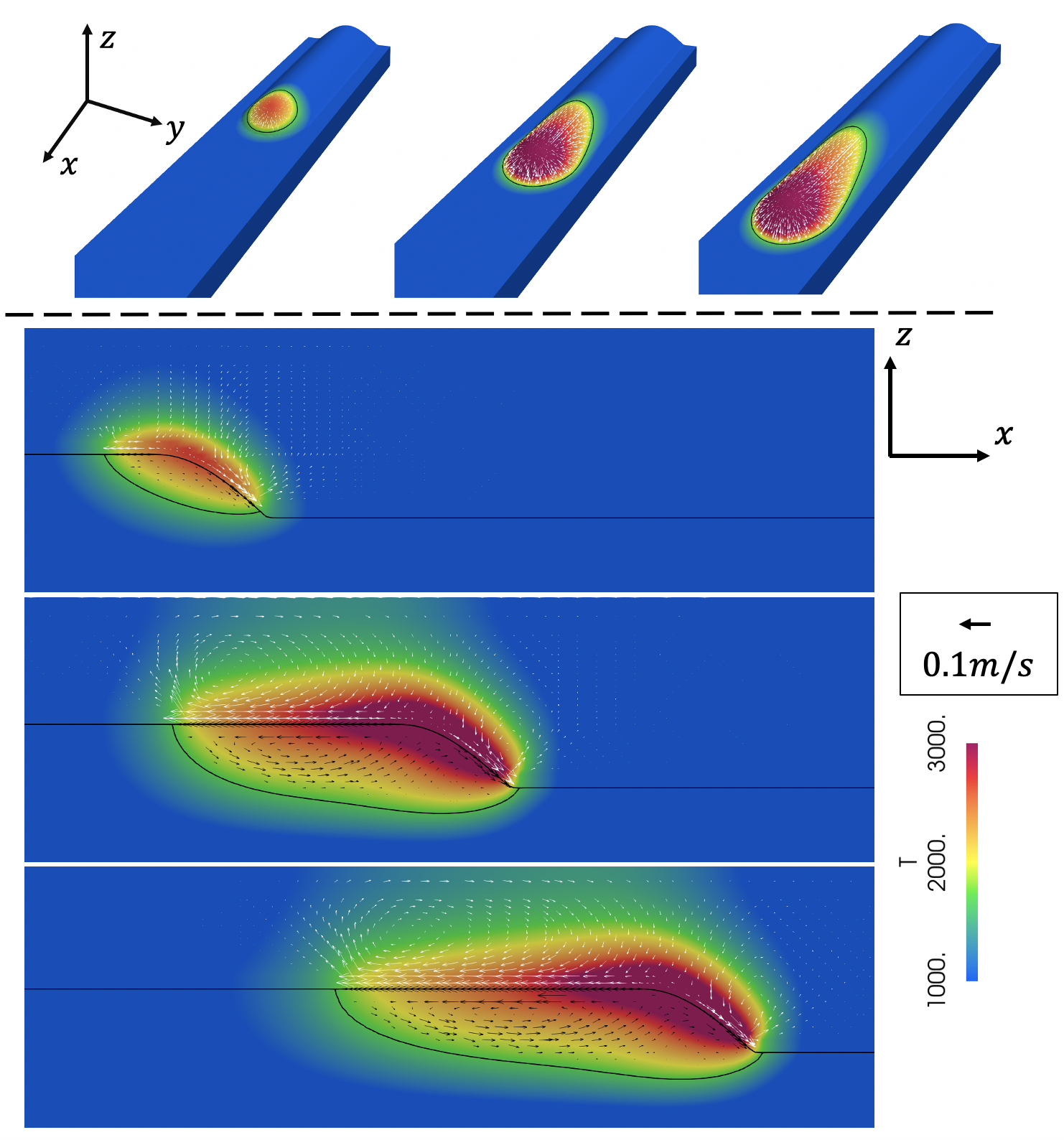}
   \caption{Velocity vectors \textcolor{black}{(unit: m/s)} and temperature \textcolor{black}{(unit: K)} contour at $t$ = 0.25 s, 0.5 s, and 0.75 s (from left to right and \textcolor{black}{from top to bottom}). \textcolor{black}{The solid line indicates the melt pool boundary and gas-metal interface. The velocity vectors are plotted in white for the gas phase and in black for the metal phase.}}
   \label{fig:ded1}
\end{figure}

\begin{figure}[!htbp]
 \centering
 \includegraphics[width=\linewidth]{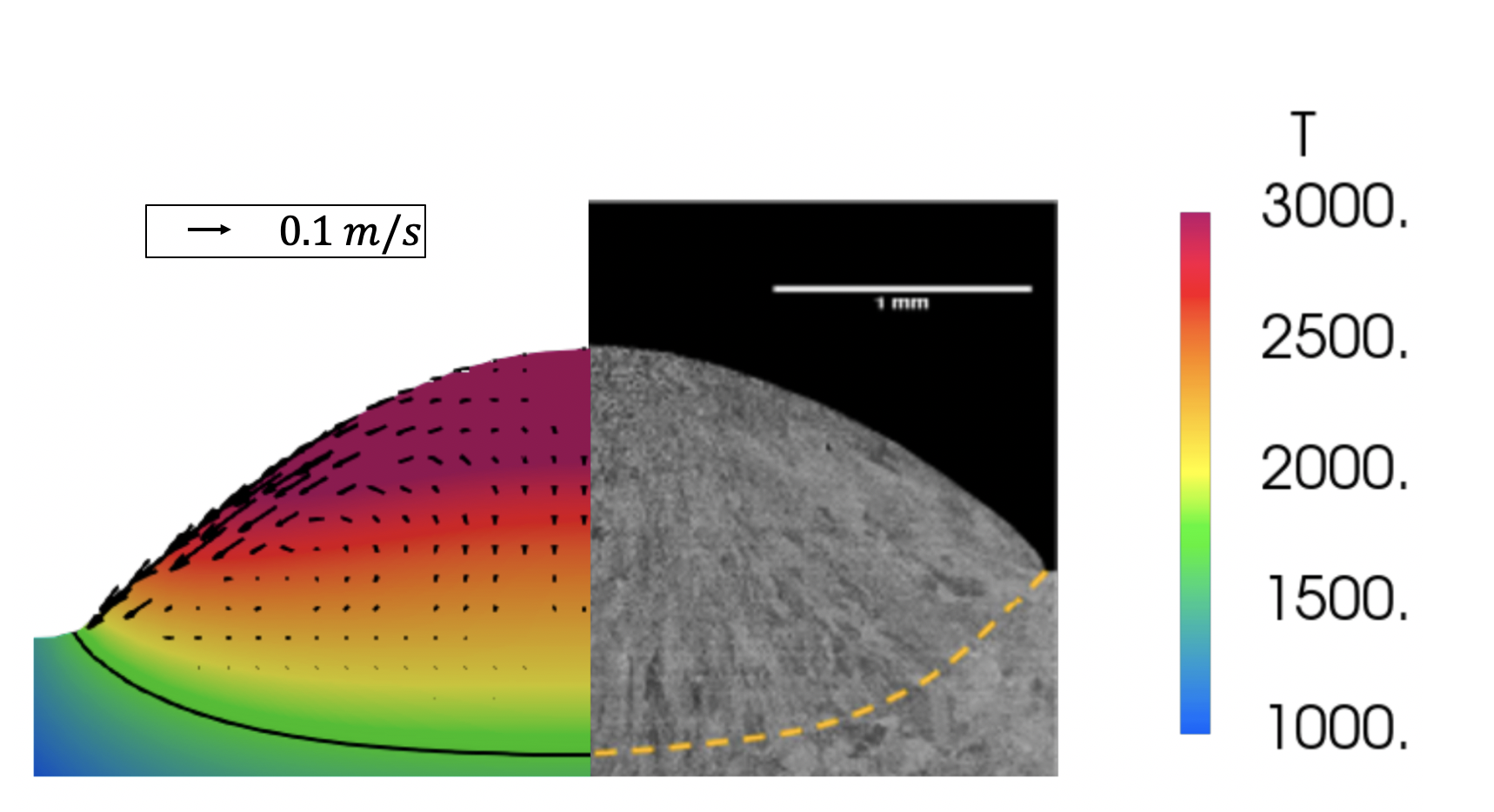}
 \caption{Cross-section of the deposit: \textcolor{black}{Temperature contour (unit: K) and velocity vectors (unit: m/s)} in the melt pool. Left: Present prediction (the solid line indicates the boundary of the melt pool). Right: Experimental image from~\cite{debroy2017digital} (the dotted yellow lines indicates the edge of the dilution region).}
 \label{fig:bubbleshape}
\end{figure}

\begin{figure}[!htbp]
 \centering
 \includegraphics[scale=0.8]{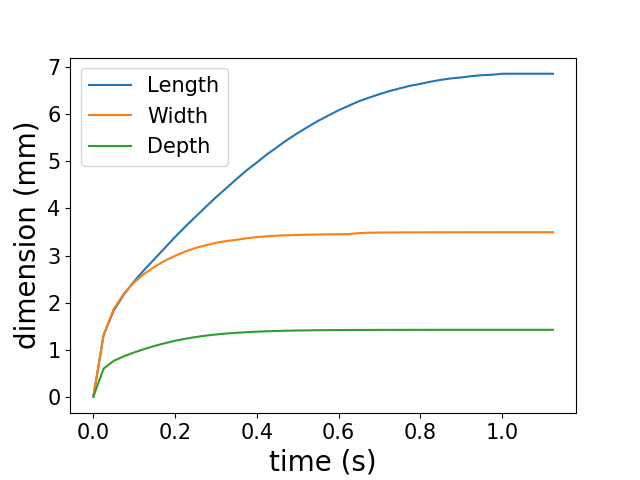}
 \caption{Time history of melt pool dimensions.}
 \label{fig:ded_geohis}
\end{figure}

Fig.~\ref{fig:ded1} shows the temperature contour, melt pool shape, and velocity vectors in \textcolor{black}{the gas} and the melt pool at $t$ = 0.25 s, 0.5 s, and 0.75 s. As the laser moves forward, the constant negative Marangoni coefficient (see Table~\ref{tab:ded_param}) drives the flow from high temperature to low temperature, leading to a long flow circulation region behind. Fig.~\ref{fig:bubbleshape} shows the built cross-section behind the laser and velocity vectors in the melt pool. The experimental image is also included for comparison. The theoretical deposit geometry model gives a very similar deposit shape to the experimental measurement. Besides, the predictive melt pool \textcolor{black}{agrees well with} the dilution region measured by the experiment. Fig.~\ref{fig:ded_geohis} shows the time history of melt pool dimension development. It takes a longer time for the length than width/depth to become stable. When the melt pool reaches the quasi-steady state (the shape does not change), the dimensions are measured and listed in Table~\ref{tab:ded_dim}. 

\begin{table}[!t]
 \caption{Material properties of SS-316L.}
 \label{tab:ded_param}
\begin{tabular}{p{4cm}p{3cm}p{4cm}}
\hline\noalign{\smallskip}
Name & Notation (units) & Value  \\
\noalign{\smallskip}\svhline\noalign{\smallskip}
   Gas density & $\rho_g$ $\textcolor{black}{\mathrm{(kg\cdot m^{-3} }})$ & 0.864 \\
   Liquid density & $\rho_l$ $\textcolor{black}{\mathrm{(kg\cdot m^{-3}) }}$ & 7800 \\
   Solid density & $\rho_s$ $\textcolor{black}{\mathrm{(kg\cdot m^{-3} }})$ & 7800 \\
   Gas heat capacity & $c_{p,g}$ $\textcolor{black}{\mathrm{(J\cdot kg^{-1} \cdot K^{-1} }})$ & 680 \\
   Liquid heat capacity & $c_{p,l}$ $\textcolor{black}{\mathrm{(J\cdot kg^{-1} \cdot K^{-1} }})$ & 769.9 \\
   Solid heat capacity & $c_{p,s}$ $\textcolor{black}{\mathrm{(J\cdot kg^{-1} \cdot K^{-1}) }}$ & $330.9+0.563T-4.015\times 10^{-4}T^2$\\
   & & $+9.465\times 10^{-8}T^3$ \\
   Gas conductivity & $k_g$ $\textcolor{black}{\mathrm{(W\cdot m^{-1}\cdot K^{-1}) }}$ & 0.028 \\
   Liquid conductivity & $k_l$ $\textcolor{black}{\mathrm{(W\cdot m^{-1}\cdot K^{-1} }})$ & 40.95 \\
   Solid conductivity & $k_s$ $\textcolor{black}{\mathrm{(W\cdot m^{-1}\cdot K^{-1}) }}$ & 11.82+0.0106T\\
   Liquidus temperature & $T_l$ $\textcolor{black}{\mathrm{(K) }}$ & 1733 \\
   Solidus temperature & $T_s$ $\textcolor{black}{\mathrm{(K) }}$ & 1693 \\
   Latent heat of fusion & $L$ $\textcolor{black}{\mathrm{(kJ\cdot kg^{-1}\cdot K }})$ & 272 \\
   Dynamics viscosity & $\mu$ $\textcolor{black}{\mathrm{(Pa\cdot s) }}$ & 0.007 \\
   Surface tension  & $\sigma$ $\textcolor{black}{\mathrm{(N\cdot m^{-1}) }}$ & 1.5 \\
   Marangoni coefficient & $\frac{\partial \sigma}{\partial T}$ $\textcolor{black}{\mathrm{(N\cdot m^{-1}\cdot K^{-1}) }}$ &  $-4\times 10^{-4}$ \\
   Ambient temperature & $T_\infty$ $\textcolor{black}{\mathrm{(K) }}$ & 300 \\
\noalign{\smallskip}\hline\noalign{\smallskip}
\end{tabular}
\end{table}

\begin{table}[!t]
 \caption{DED parameters.}
 \label{tab:dedproc_param}
\begin{tabular}{p{4cm}p{3cm}p{4cm}}
\hline\noalign{\smallskip}
Name & Notation (units) & Value  \\
\noalign{\smallskip}\svhline\noalign{\smallskip}
   Laser power & $Q$ $\textcolor{black}{\mathrm{(W) }}$ & $2500$ \\
   Laser moving speed &$V_s$ $\textcolor{black}{\mathrm{(m\cdot s^{-1}) }}$& $0.0106$\\
   Laser radius & $r_b$ $\textcolor{black}{\mathrm{(m) }}$ & $0.002$ \\
   distribution factor & $d$ $\textcolor{black}{(-)}$& $2.0$ \\
   Powder flow rate & $\dot{m}$ $\textcolor{black}{\mathrm{(kg\cdot s^{-1}) }}$ & $0.25\times 10^{-3}$\\
\noalign{\smallskip}\hline\noalign{\smallskip}
\end{tabular}
\end{table}

\begin{table}[!t]
 \caption{Quasi-steady melt pool dimensions.}
 \label{tab:ded_dim}
\begin{tabular}{p{4cm}p{3cm}p{4cm}}
\hline\noalign{\smallskip}
Length ($\mathrm{mm}$) & Width ($\mathrm{mm}$) & Depth ($\mathrm{mm}$) \\
\noalign{\smallskip}\svhline\noalign{\smallskip}
   6.80 & 3.40 & 1.35 \\
\noalign{\smallskip}\hline\noalign{\smallskip}
\end{tabular}
\end{table}

\begin{figure}[!htbp]
 \centering
 \includegraphics[scale=0.8]{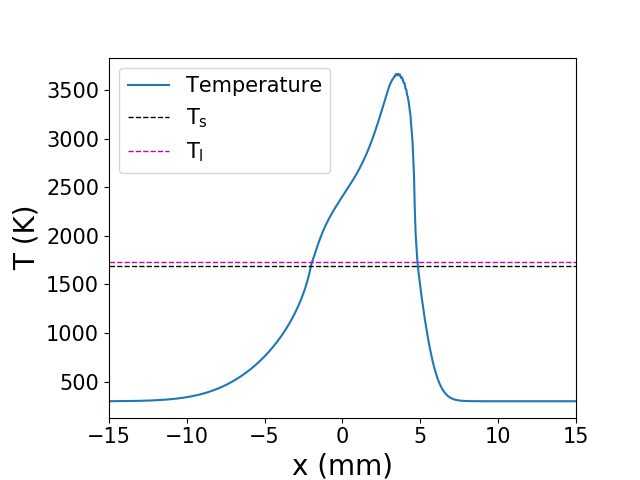}
 \caption{Temperature profile along the centerline of the deposit at the quasi-steady state.}
 \label{fig:centerline}
\end{figure}

Cooling rate is a vital variable in metal AM, which has an unmatched effect on micro-structure evolution~\cite{gan2019nist, debroy2017digital}, such as dendrite growth. Previous research indicated that the mechanical strength of additive manufactured parts is closely related to cooling rate. Fig.~\ref{fig:centerline} shows the temperature profile at the quasi-steady state along the centerline of the deposit. In this paper, cooling rate is calculated at the centerline of the deposit by $K_v = \frac{T_l-T_s}{t_v}$, where $t_v$ is the time of the cooling process takes from liquidus temperature to solidus temperature. We employ the theoretical model from~\cite{yin2010dendrite} to evaluate the effect of cooling rate on mechanical properties. Firstly, the average magnitude of secondary dendrite arm spacing (SDAS), $\lambda$, is computed as
\begin{align}
\lambda=50(K_v)^{-0.4}
\end{align}
The unit of $\lambda$ is $\textcolor{black}{\mathrm{\mu m}}$. Moreover, the average hardness $H_\nu$ related to yield strength of SS-316L takes the following form~\cite{debroy2014activation}
\begin{align}
H_\nu=3\sigma_y(0.1)^{-\frac{1}{4}}
\end{align}
where $\sigma_y$ is the yield strength, which is a also function of average magnitude of SDAS, defined as
\begin{align}
\sigma_y=\sigma_0+\frac{K_y}{\sqrt{\lambda}}
\end{align}
where $\sigma_0$ and $K_y$ are material-dependent coefficients, whose values are $240$ $\textcolor{black}{\mathrm{MPa}}$ and $279$ $\textcolor{black}{\mathrm{MPa}}\cdot\textcolor{black}{\mathrm{\mu m^{\frac{1}{2}}}}$ for SS-316L, respectively~\cite{kashyap1995am}. Based these models, we \textcolor{black}{list the predicted} cooling $K_v$, average magnitude of SDAS $\lambda$, and averaged hardness $H_v$ in~Table~\ref{tab:ded_hv}, which shows good agreement with the experimental and simulation data obtained from~\cite{debroy2017digital}.

\begin{table}[!t]
 \caption{Cooling rate $K_v$, average magnitude of SDAS $\lambda$, and average hardness $H_\nu$.}
 \label{tab:ded_hv}
\begin{tabular}{p{3cm}p{3cm}p{2cm}p{2cm}}
\hline\noalign{\smallskip}
{}& Cooling rate $\textcolor{black}{\mathrm{(K\cdot s^{-1})}}$ & SDAS $\textcolor{black}{\mathrm{(\mu m)}}$ & Hardness $\textcolor{black}{\mathrm{(MPa)}}$ \\
\noalign{\smallskip}\svhline\noalign{\smallskip}
 Present & 937 & 3.24 & 2107.5 \\
 Simulation~\cite{debroy2017digital}  & 608 & 3.85& 2039.0\\
 Experiment~\cite{debroy2017digital}  & - &3.27$\pm$0.65 & 2014.5$\pm$44.5\\
\noalign{\smallskip}\hline\noalign{\smallskip}
\end{tabular}
\end{table}

\section{Conclusion}\label{sec:conclusion}

This book chapter summarizes recent method developments from the authors for simulating thermal multi-phase flows in metal AM processes. A mixed interface-capturing/interface-tracking approach and a new deposit geometry model were presented. The development aims to address the limitations of isolated interface-capturing and interface-tracking methods on metal AM process applications. The mixed formulation takes full advantage of both interface-capturing and interface-tracking methods to better handle the gas-metal interface, where AM physics, such as phase transitions and laser-material interaction, mainly occurs. Four major contributions of the paper are:

\begin{enumerate}
   \item  A simple computational geometry-based re-initialization technique, which maintains excellent signed distance property on unstructured meshes, re-constructs an explicit representation of gas-metal interface, and facilitates the treatment of the multiple laser reflections during keyhole evolution in AM processes; 
   \item A fully coupled VMS formulation for thermal multi-phase governing equations, including Navier-stokes, level set convection, and thermodynamics with melting, solidification, evaporation, and interfacial force models; 
   \item A three-level recursive preconditioning technique to enhance the robustness of linear solvers.
\item A physics-based and non-empirical deposit geometry model based on an energy minimization with a mass conservation constraint.
\end{enumerate}

 We demonstrate the proposed formulation's accuracy and modeling capabilities on a set of numerical examples, including the most recent metal AM experiments performed by Argonne National Lab. The results show the great potential of the formulation in the broad application in advanced manufacturing.  Some parts of the formulation can be polished, which will be addressed in the authors' subsequent development. Firstly, we will extend the geometry-based re-initialization to other spatial discretizations, such as non-uniform rational b-splines (NURBS) in IGA. Secondly, in this paper, a large portion of the mesh is pre-refined along the laser track, which can be inefficient. To this end, adaptive mesh schemes will be incorporated into the thermal multi-phase flow formulation. We will also develop efficient computational geometry algorithms to deploy the geometry-based re-initialization to the adaptive meshes.



\bibliographystyle{unsrtnat}
\bibliographystyle{plain}
\bibliographystyle{elsarticle-num}
%
%
%

\end{document}